\documentclass[twocolumn]{aastex63}

\usepackage{subfigure}
\usepackage{url}
\usepackage{hyperref}
\usepackage{epsfig}
\usepackage{graphicx}
\usepackage{sidecap}
\usepackage{hanging}
\usepackage{color}
\usepackage{multirow}
\usepackage{enumerate}
\usepackage{natbib}
\usepackage{amssymb}
\usepackage{amsmath}
\usepackage[bottom]{footmisc}

\newcommand{\sci}{Science}
\newcommand{\jatis}{JATIS}

\newcommand{\Kepler}{{Kepler}}
\newcommand{\TESS}{{TESS}}

\newcommand{\Gaia}{{Gaia}}

\providecommand{\e}[1]{\ensuremath{\, \times \, 10^{#1}}}

\newcommand{\kic}{{KIC~5951458}}
\newcommand{\kicb}{{KIC~5951458~b}}
\submitjournal{AJ}

\shorttitle{The Single Occultation of KIC 5951458}
\shortauthors{Dalba et al.}

\begin{document}

\title{Multiple Explanations for the Single Transit of KIC~5951458 Based on Radial Velocity Measurements Extracted with a Novel Matched-template Technique\footnote{Some of the data presented herein were obtained at the W. M. Keck Observatory, which is operated as a scientific partnership among the California Institute of Technology, the University of California, and the National Aeronautics and Space Administration. The Observatory was made possible by the generous financial support of the W. M. Keck Foundation.}}

\correspondingauthor{Paul A. Dalba}
\email{pdalba@ucr.edu}

%%%%%%%%%%%%%%%%%%%%%%%%%%%%%%%%%%%%%%%%%%%%%%%%%%%%%%%%%%%%%%%%%%%%%%%%%%

\author[0000-0002-4297-5506]{Paul A.\ Dalba}
\altaffiliation{NSF Astronomy and Astrophysics Postdoctoral Fellow.}
\affiliation{Department of Earth and Planetary Sciences, University of California Riverside, 900 University Ave., Riverside, CA 92521, USA}

\author[0000-0003-3504-5316]{Benjamin Fulton}
\affiliation{NASA Exoplanet Science Institute/Caltech-IPAC, MC 314-6, 1200 E California Blvd., Pasadena, CA 91125, USA}

\author[0000-0002-0531-1073]{Howard Isaacson}
\affiliation{{Department of Astronomy,  University of California Berkeley, Berkeley, CA 94720, USA}}
\affiliation{Centre for Astrophysics, University of Southern Queensland, Toowoomba, QLD, Australia}

\author[0000-0002-7084-0529]{Stephen R.\ Kane}
\affiliation{Department of Earth and Planetary Sciences, University of California Riverside, 900 University Ave, Riverside, CA 92521, USA}

\author[0000-0001-8638-0320]{Andrew W.\ Howard}
\affiliation{Department of Astronomy, California Institute of Technology, Pasadena, CA 91125, USA}

%%%%%%%%%%%%%%%%%%%%%%%%%%%%%%%%%%%%%%%%%%%%%%%%%%%%%%%%%%%%%%%%%%%%%%%%%%

\begin{abstract}
Planetary systems that show single-transit events are a critical pathway to increasing the yield of long-period exoplanets from transit surveys. From the primary Kepler mission, KIC~5951458~b (Kepler-456b) was thought to be a single-transit giant planet with an orbital period of 1310~days. However, radial velocity (RV) observations of KIC~5951458 from the HIRES instrument on the Keck telescope suggest that the system is far more complicated. To extract precise RVs for this $V\approx13$ star, we develop a novel matched-template technique that takes advantage of a broad library of template spectra acquired with HIRES. We validate this technique and measure its noise floor to be 4--8~m~s$^{-1}$ (in addition to internal RV error) for most stars that would be targeted for precision RVs. For KIC~5951458, we detect a long-term RV trend that suggests the existence of a stellar companion with an orbital period greater than a few thousand days. We also detect an additional signal in the RVs that is possibly caused by a planetary or brown dwarf companion with mass in the range of 0.6--82~$M_{\rm Jup}$ and orbital period below a few thousand days. Curiously, from just the data on hand, it is not possible to determine which object caused the single ``transit'' event. We demonstrate how a modest set of RVs allows us to update the properties of this unusual system and predict the optimal timing for future observations.
\end{abstract}


%%%%%%%%%%%%%%%%%%%%%%%%%%%%%%%%%%%%%%%%%%%%%%%%%%%%%%%%%%%%%%%%%%%%%%%%%%
 
\section{Introduction} \label{sec:intro}

The vast majority of known transiting exoplanets whip around their host stars on orbits of 10~days or less. The detection of transiting exoplanets with short orbital periods results from the twofold bias of the transit method. First, the geometric probability of observing an exoplanet transit scales inversely with the physical separation between the planet and the star. Second, the finite observational baseline reduces the probability of detecting exoplanets that have relatively long orbital periods. Considering both biases, the best means of extending the range of orbital periods of known transiting exoplanets is to observe many stars for as long as possible. 

This was the approach of the primary \Kepler\ mission, which achieved a nearly continuous, 4 yr observational baseline for over 150,000~stars \citep{Borucki2010}. This baseline is so far unrivaled among transit surveys, as is the sample of long-period transiting exoplanets identified in the \Kepler\ data set. Long-period exoplanets are more readily discovered through radial velocity (RV) measurements of their host stars. However, the fortuitous alignment that causes a transit also enables analyses that simply cannot be conducted for most RV-detected exoplanets. This means that the long-period sample of transiting exoplanets \citep[e.g.,][]{Kawahara2019} is small, but each one is extraordinarily valuable.  

Among these exoplanets is \kicb\ (also known as Kepler-456b), a supposed $\sim$6.6~Earth-radius planet candidate that was only observed to transit once during the \Kepler\ primary mission \citep{Wang2015,Kawahara2019}. The two previous efforts to characterize this planet candidate measured an orbital period in the range of 1167.6--13721.9~days \citep{Wang2015} and 1600$^{+1100}_{-400}$~days \citep{Kawahara2019}. Neither estimate is precise owing to the lack of additional transits. \kicb\ was not categorized as a \Kepler\ object of interest (KOI) by the \Kepler\ pipeline \citep[e.g.,][]{Thompson2018}, but subsequent analysis led to its statistical validation \citep{Wang2015}. Statistical validation of exoplanets is a common technique that typically involves ruling out enough of the false-positive scenario parameter space to claim the planetary nature to some probability threshold \citep[e.g.,][]{Barclay2013,Crossfield2016,Morton2016}. In the case of \kicb, the statistical validation process produced a planet probability of 99.8\% \citep{Wang2015}. Ideally, validated exoplanets like \kicb\ would also receive sufficient follow-up observation to uniquely measure their mass, which more clearly determines their nature. However, given the ever-growing number of candidates from planet discovery efforts such as \Kepler, K2 \citep{Howell2014}, and the Transiting Exoplanet Survey Satellite \citep[\TESS;][]{Ricker2015}, statistical validation techniques are becoming increasingly common. 

In this work, we explore the validated exoplanet \kicb\ by acquiring a small number of RV observations of its host star. We do not attempt to conduct a full characterization of this system; such an effort will not be possible for many single-transit candidate exoplanets that have been (or will be) discovered by \Kepler\ or \TESS. For \Kepler\ single-transit detections, baselines of several years are likely required to cover a full orbit. Additionally, the faintness of typical \Kepler\ host stars severely limits the list of facilities capable of making precise RV measurements. In the case of \TESS\ single-transit detections, time baselines are shorter and host stars are brighter \citep[e.g.,][]{Dalba2020a,Eisner2020,Gill2020b}, and the number of detections is expected to be much higher than for \Kepler\ \citep[e.g.,][]{Villanueva2019}. As a result, follow-up resources must be used strategically to maximize the science return from single-transit detections. For \kic, we collect only a modest set of RVs that we then combine with archival data to identify inaccuracies in the currently published properties of \kicb. We then demonstrate how a careful consideration of possible scenarios for the nature of the \kic\ system enables informed predictions for future observations.

In Section \ref{sec:archive}, we identify and process archival observations of \kic, which include photometry from all quarters of the \Kepler\ primary mission and adaptive optics (AO) imaging. In Sections \ref{sec:spec} and \ref{sec:match_temp}, we present RV observations of \kic\ from the High Resolution Echelle Spectrometer (HIRES) at the 10 m Keck I telescope. We describe a new method of extracting precision RV measurements that involves matching \kic\ to a star in the HIRES spectral library that already has a spectral template. We validate this method and demonstrate its benefit for faint host stars like those observed by \Kepler. In Section \ref{sec:rj}, we employ the rejection sampling package \texttt{The Joker} \citep{PriceWhelan2017} to model the RVs. We find that \kic\ likely hosts both a stellar companion and a giant planetary or brown dwarf companion, but it is not clear which one caused the occultation event observed by \Kepler. In Section \ref{sec:discussion}, we discuss the various scenarios for the nature of the \kic\ system in the context of other known exoplanets and binary star systems. We also place our efforts for this system in the context of similar ongoing research, mostly related to the \TESS\ mission. Finally, in section \ref{sec:summary}, we summarize our findings for the remarkable \kic\ system.

\begin{figure*}
  \begin{center}
    \begin{tabular}{ccc}
      \includegraphics[width=5.3cm]{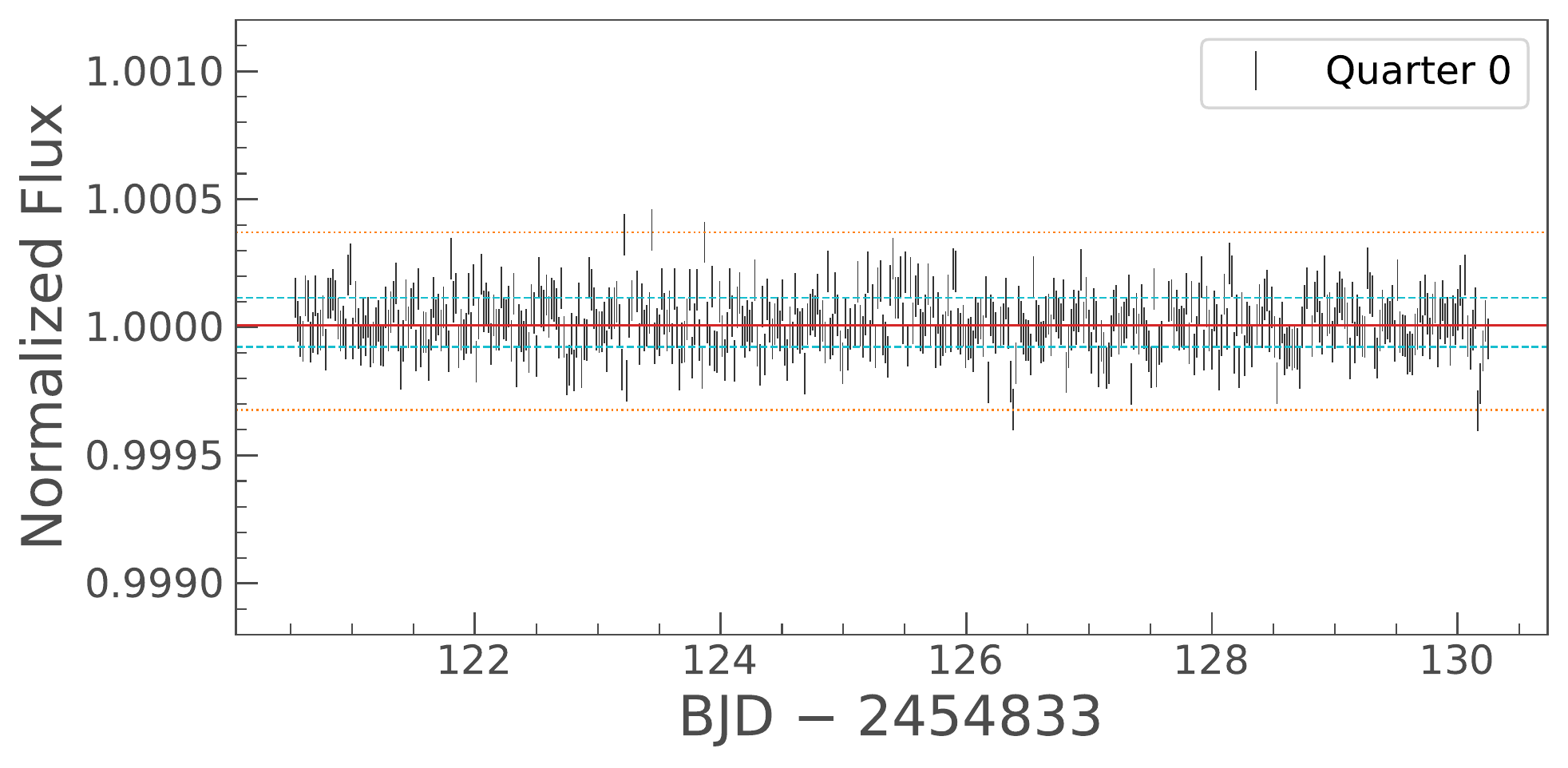} &
      \includegraphics[width=5.3cm]{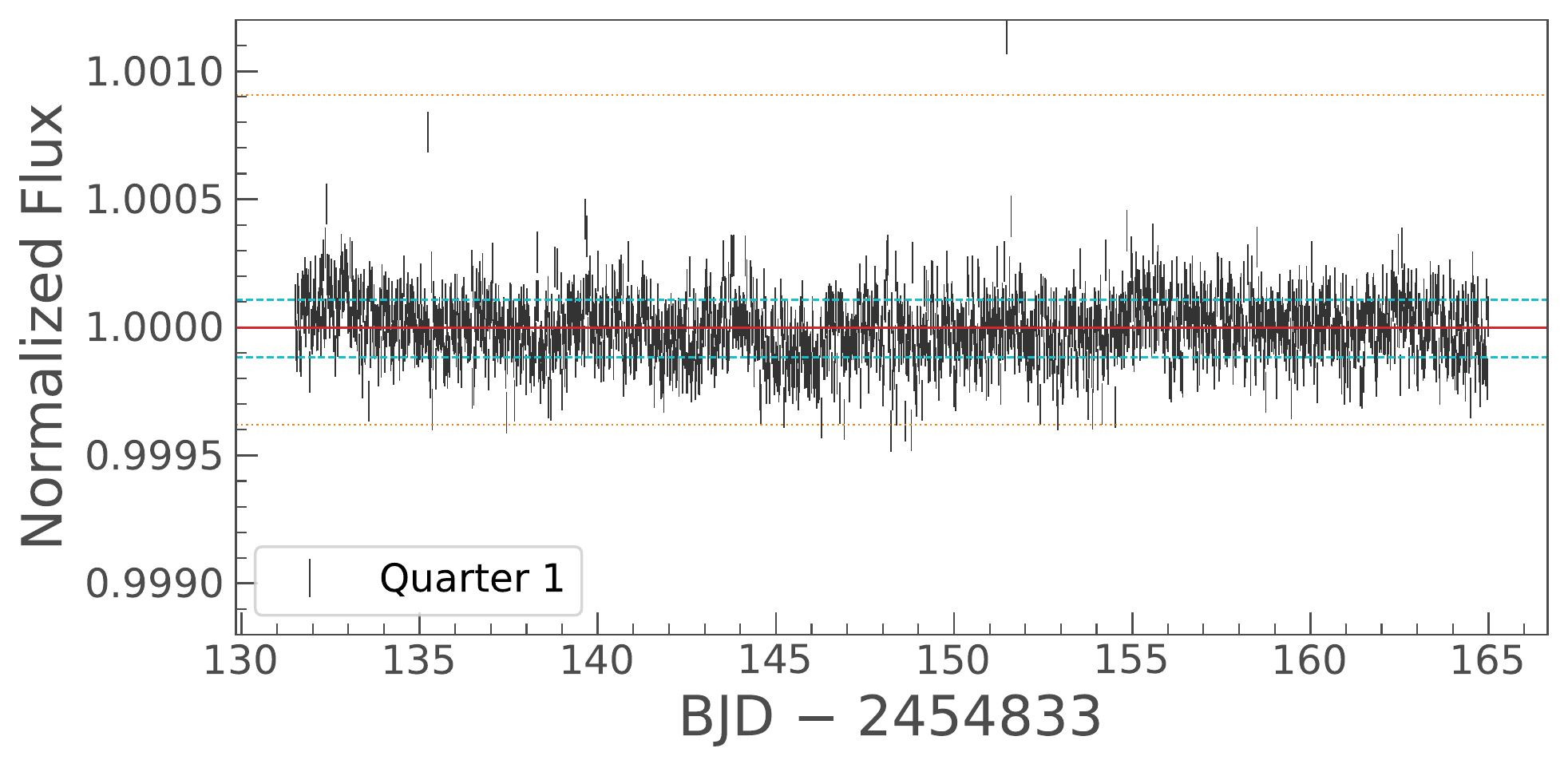} &
      \includegraphics[width=5.3cm]{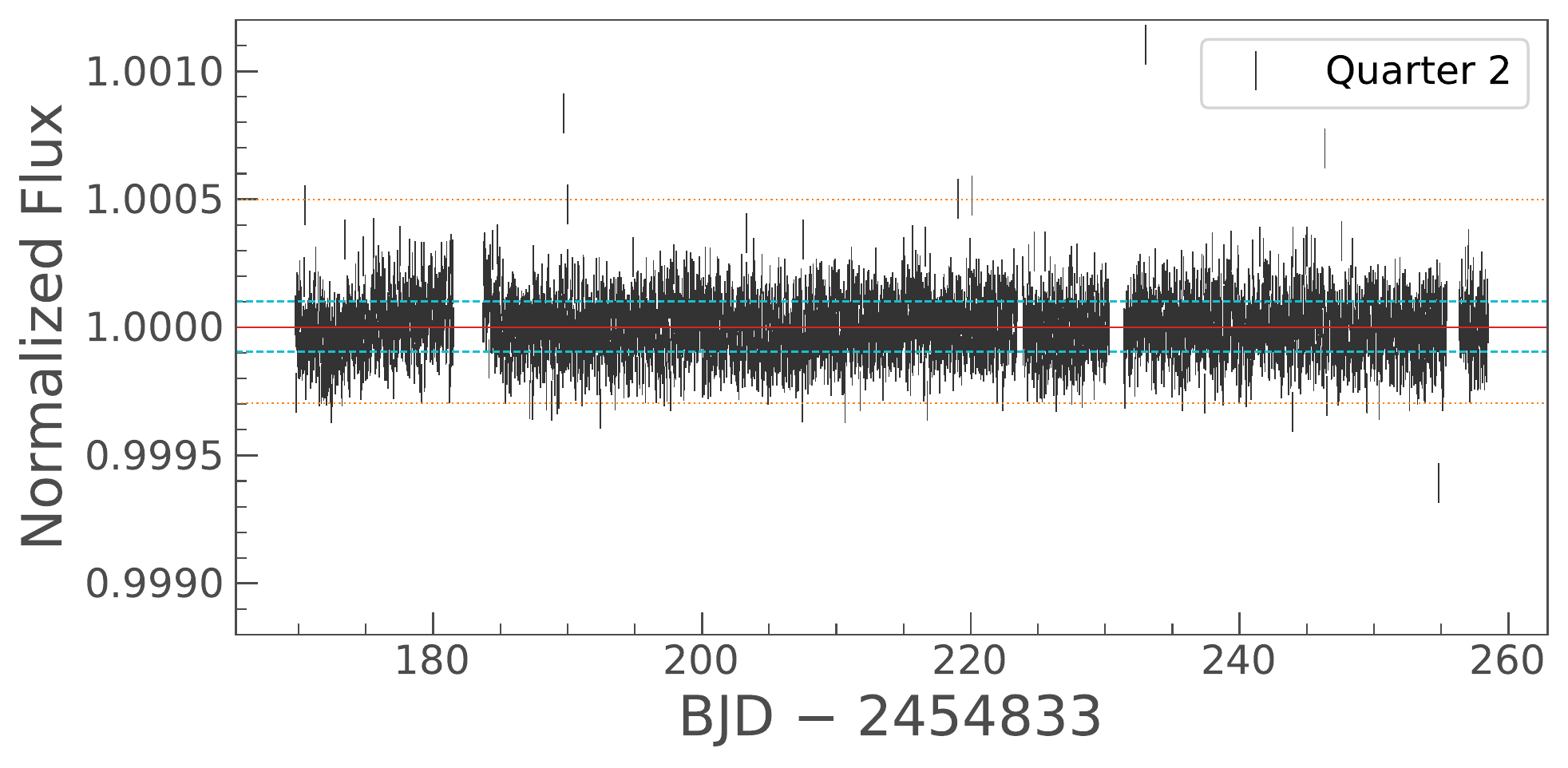} \\
      \includegraphics[width=5.3cm]{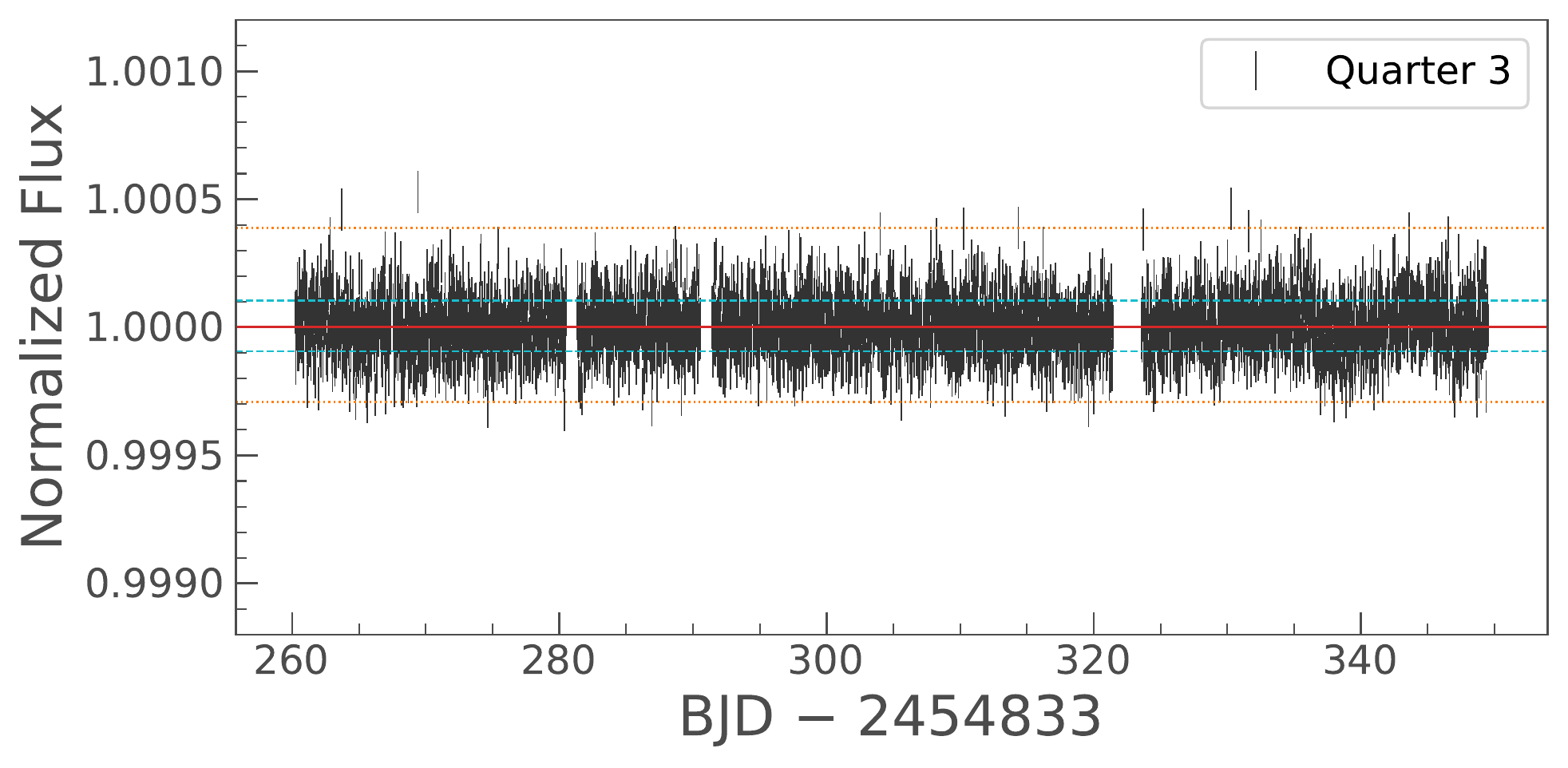} &
      \includegraphics[width=5.3cm]{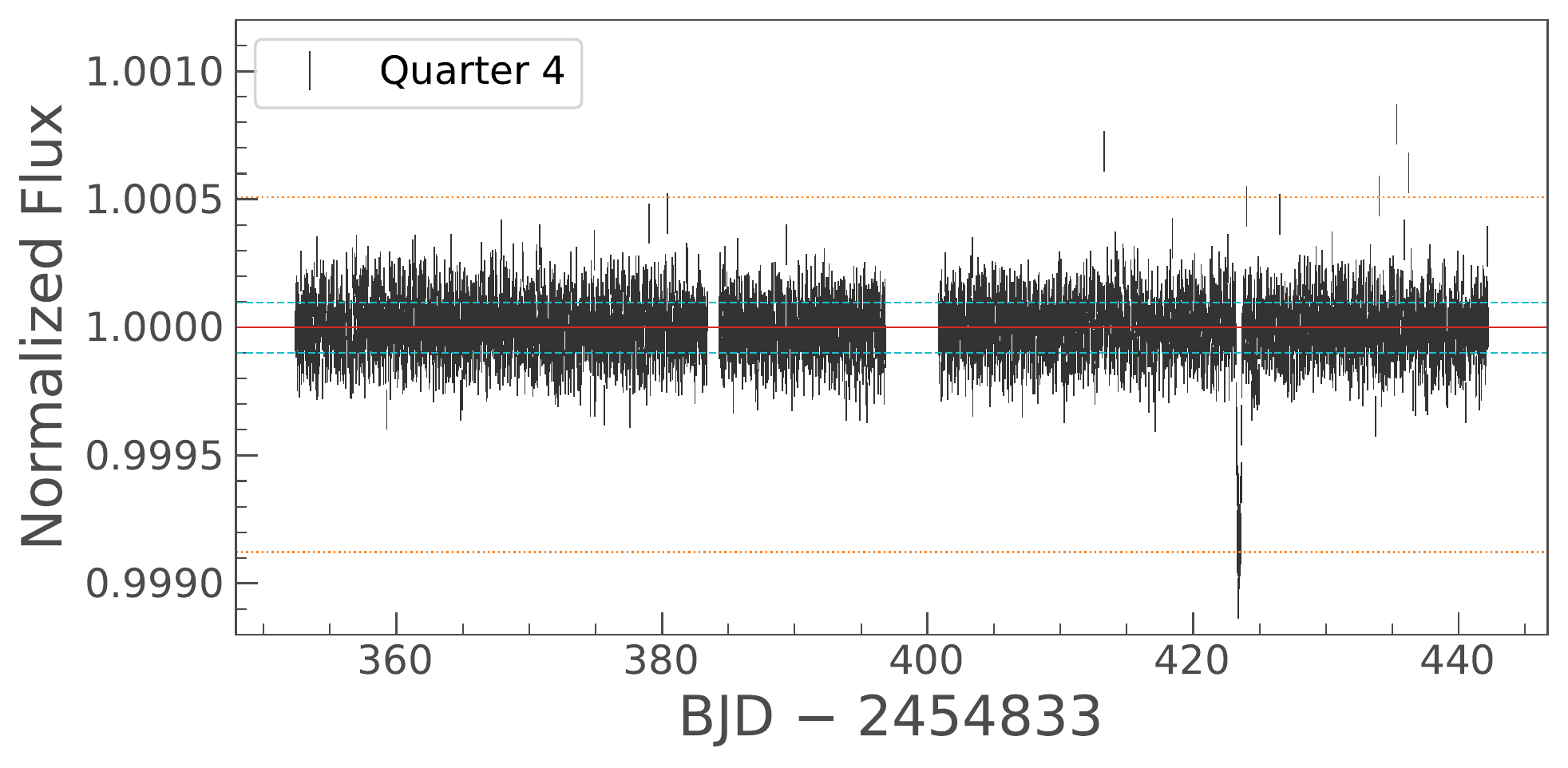} &
      \includegraphics[width=5.3cm]{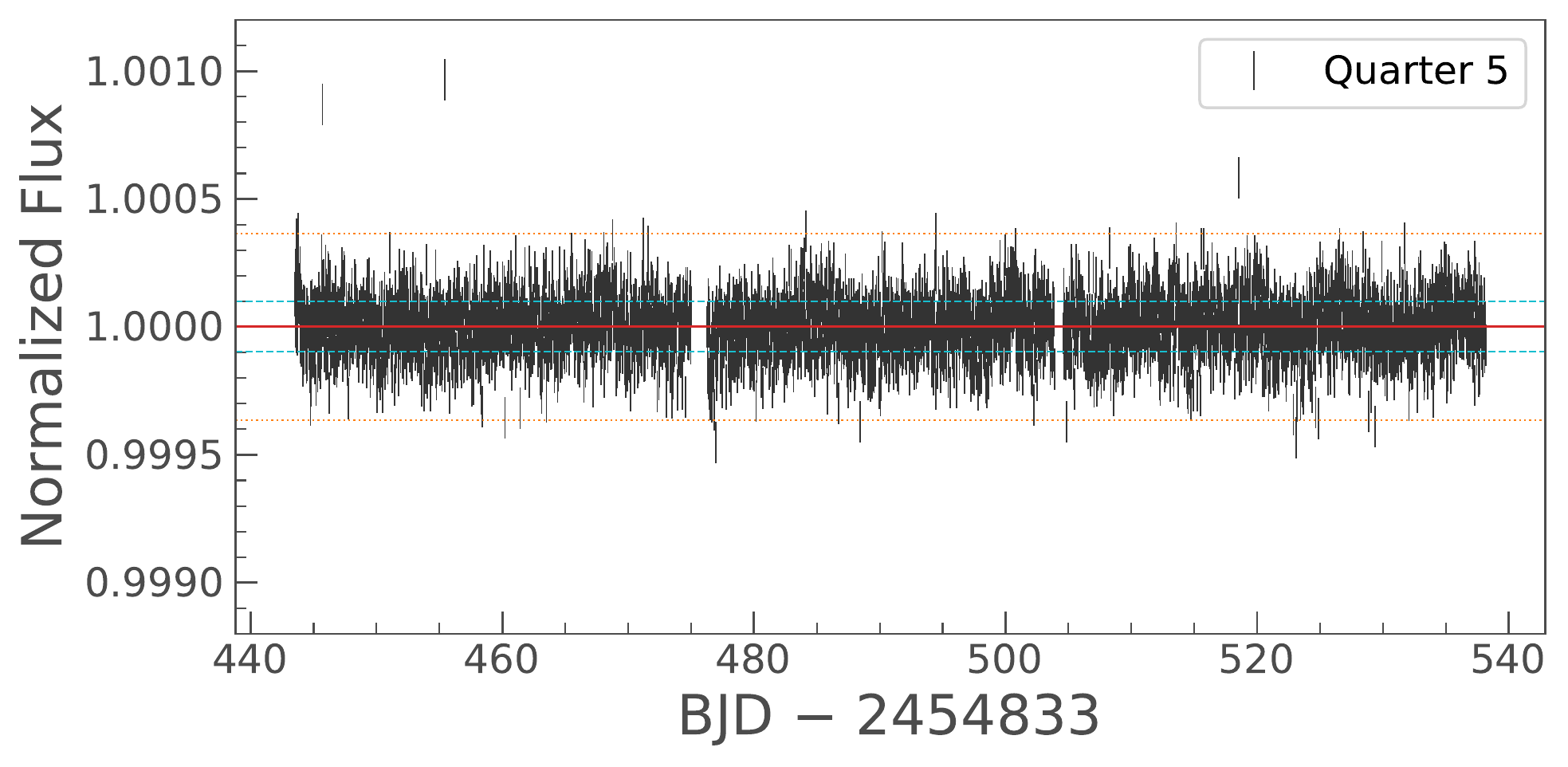} \\
      \includegraphics[width=5.3cm]{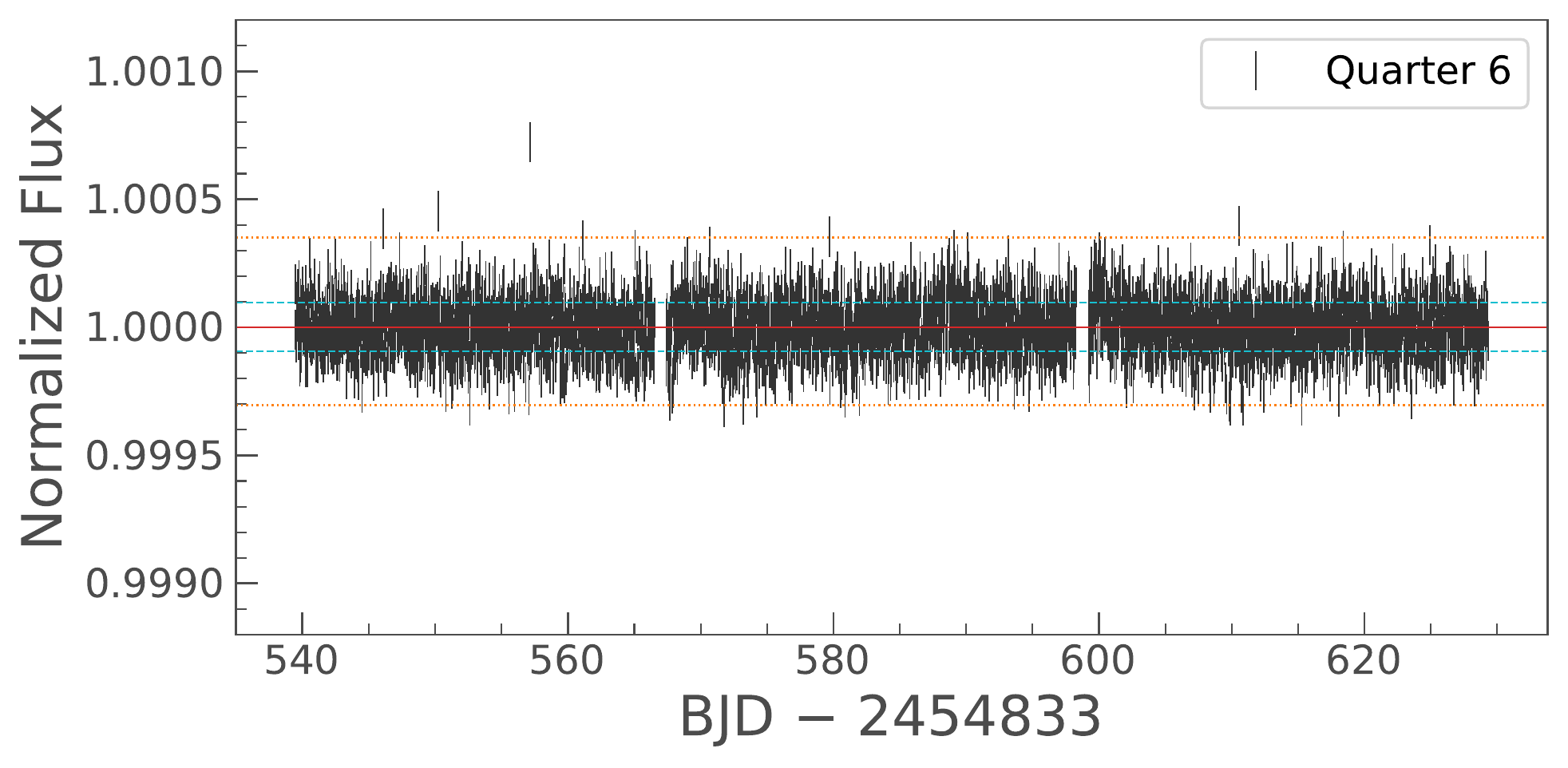} &
      \includegraphics[width=5.3cm]{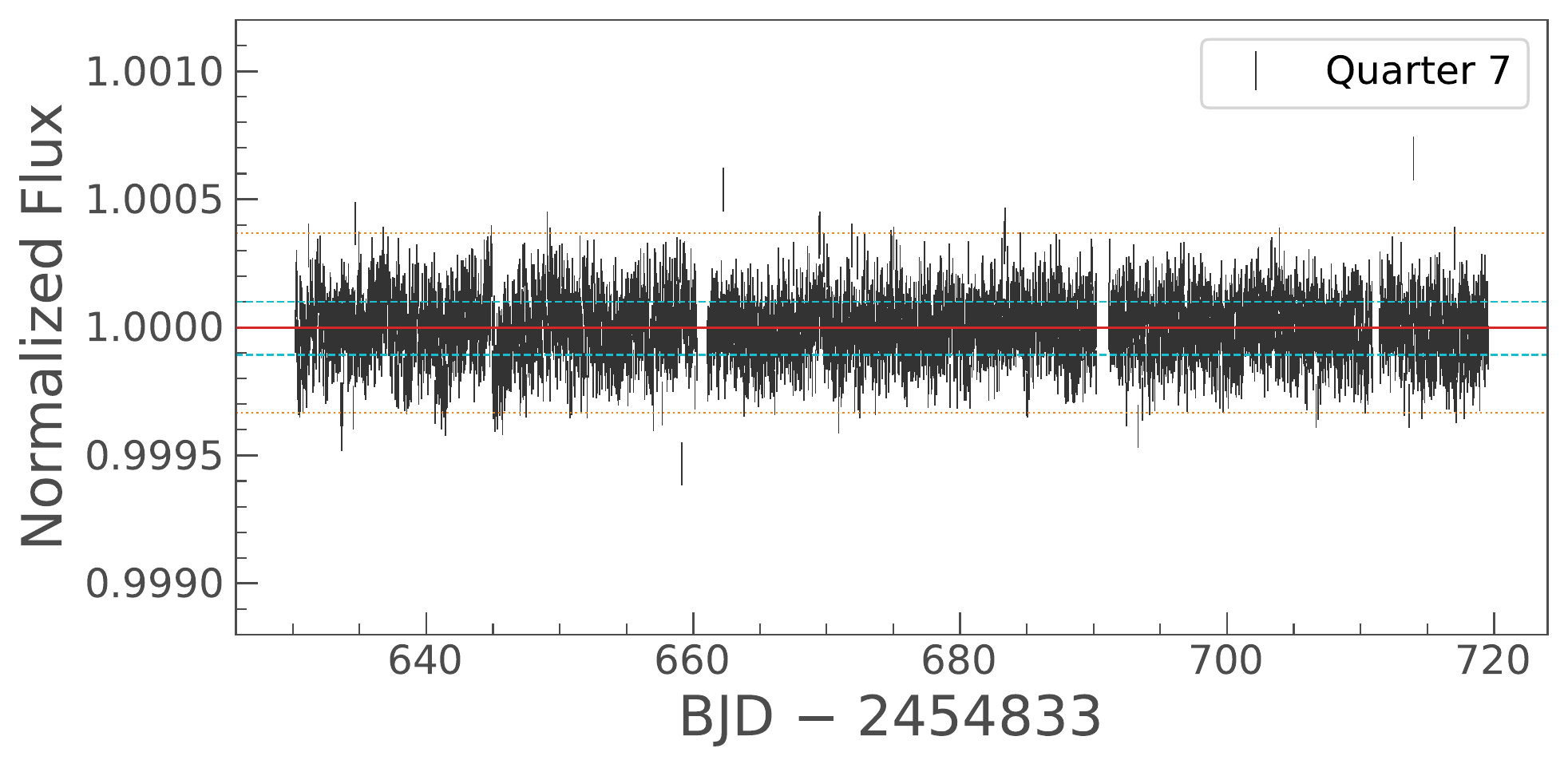} &
      \includegraphics[width=5.3cm]{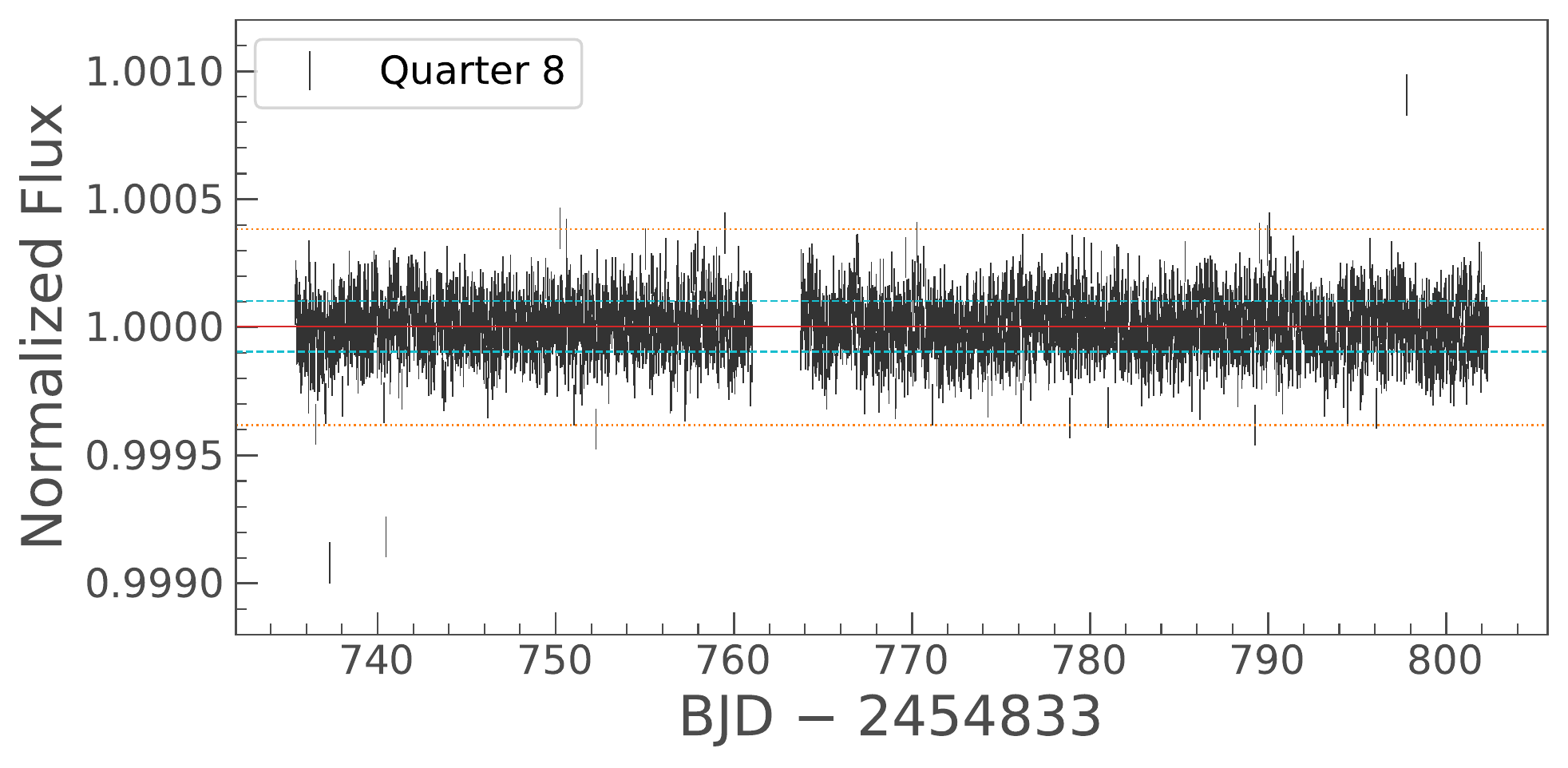} \\
      \includegraphics[width=5.3cm]{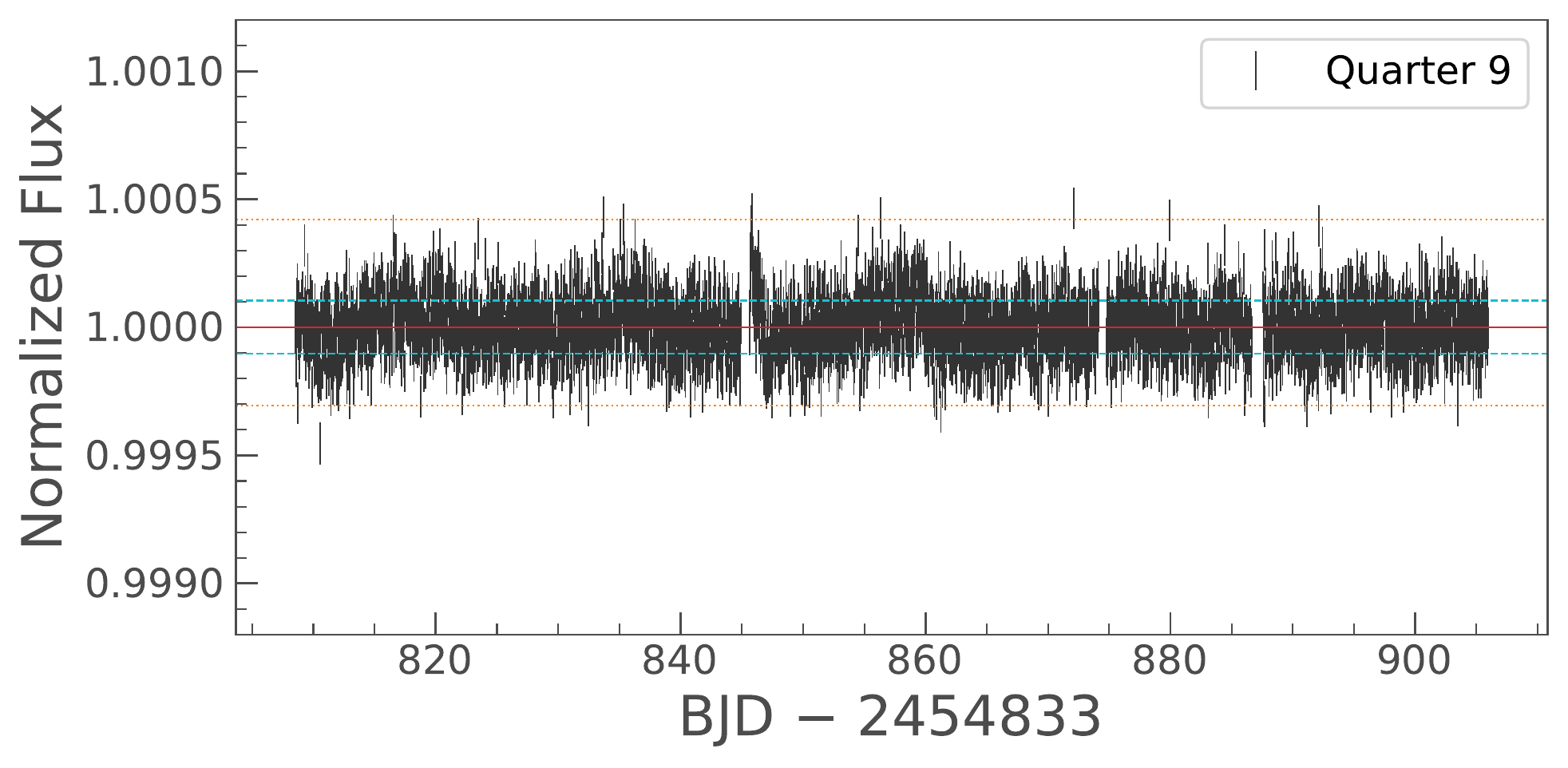} &
      \includegraphics[width=5.3cm]{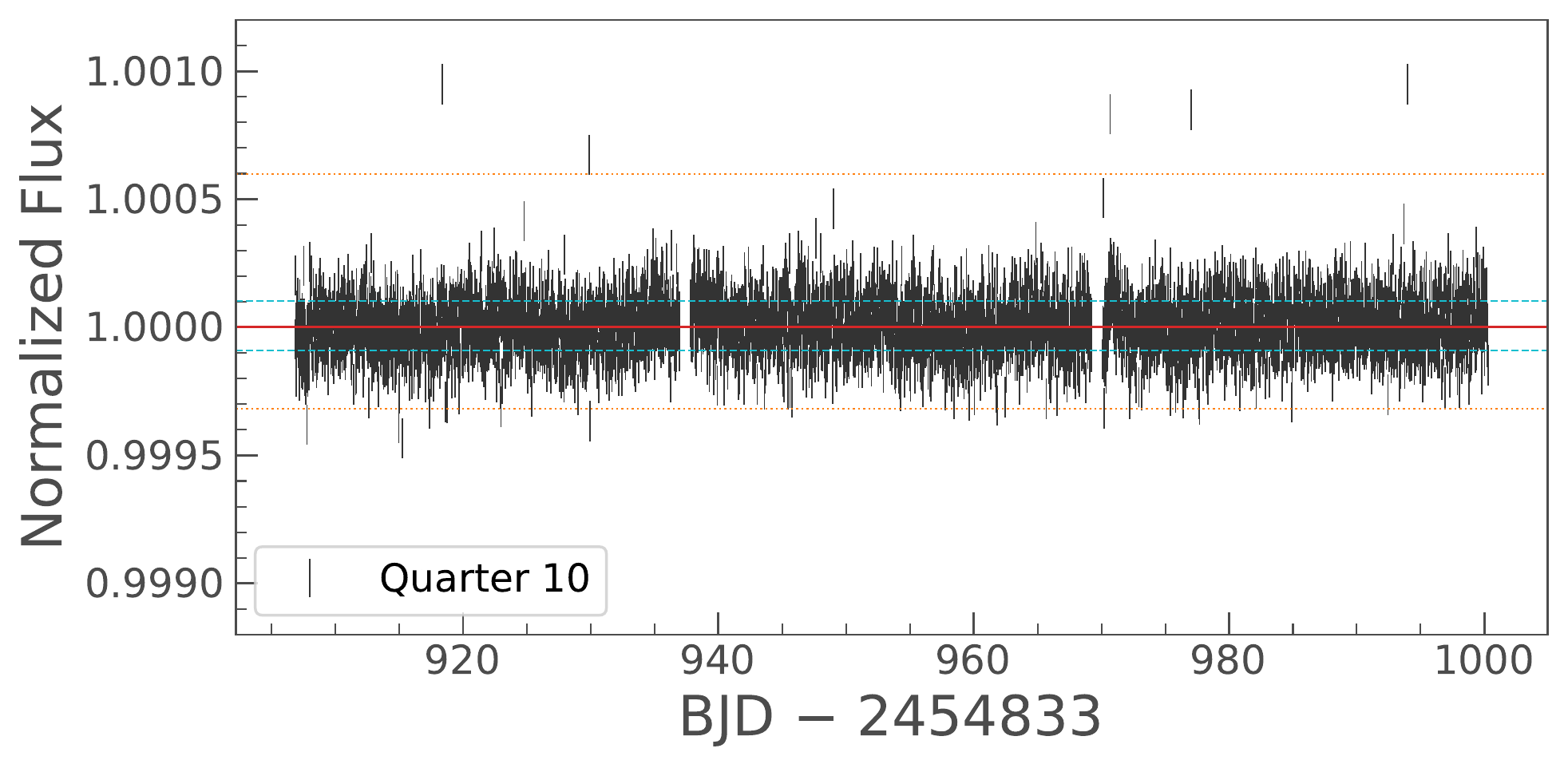} &
      \includegraphics[width=5.3cm]{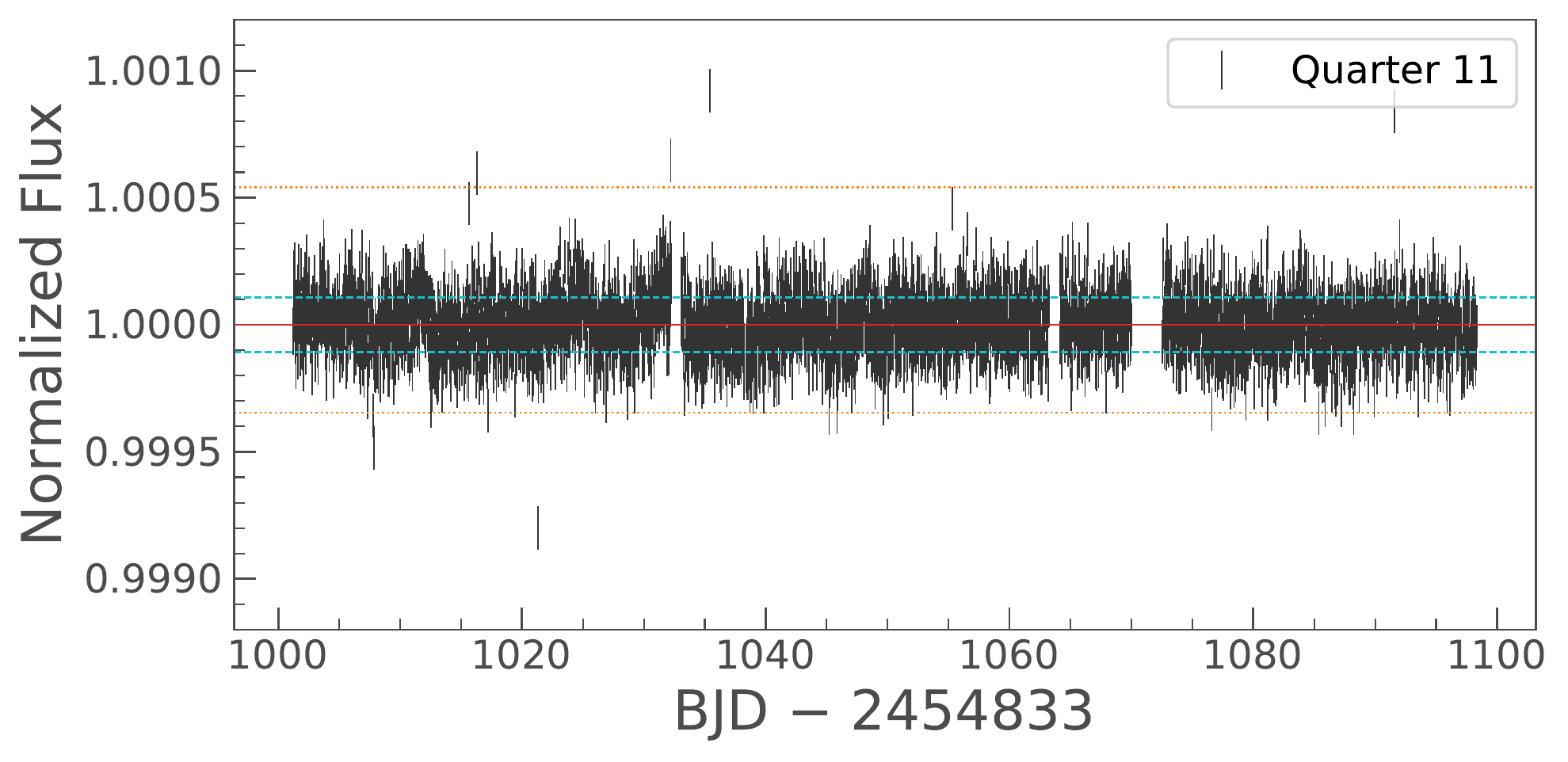} \\
      \includegraphics[width=5.3cm]{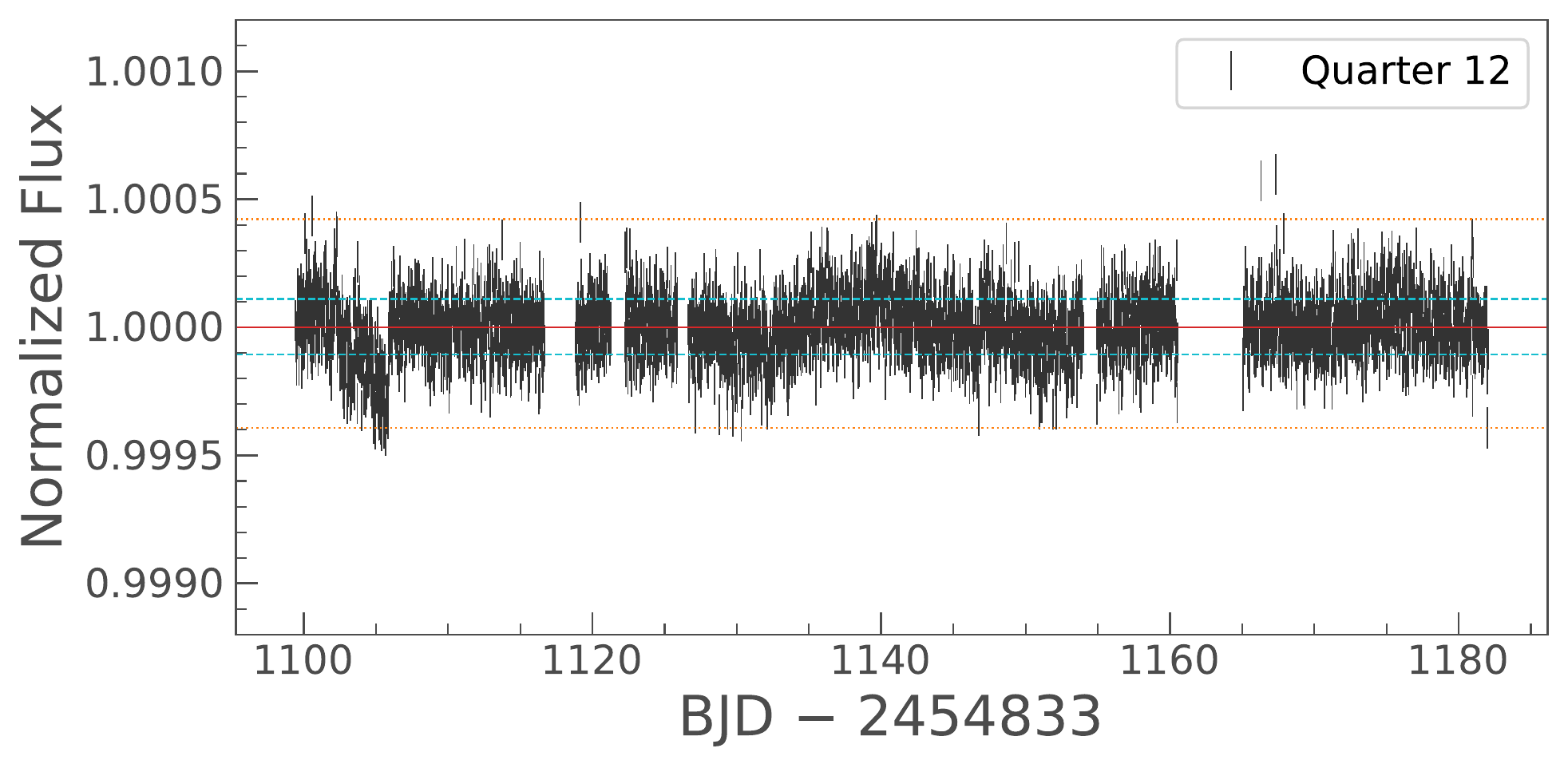} &
      \includegraphics[width=5.3cm]{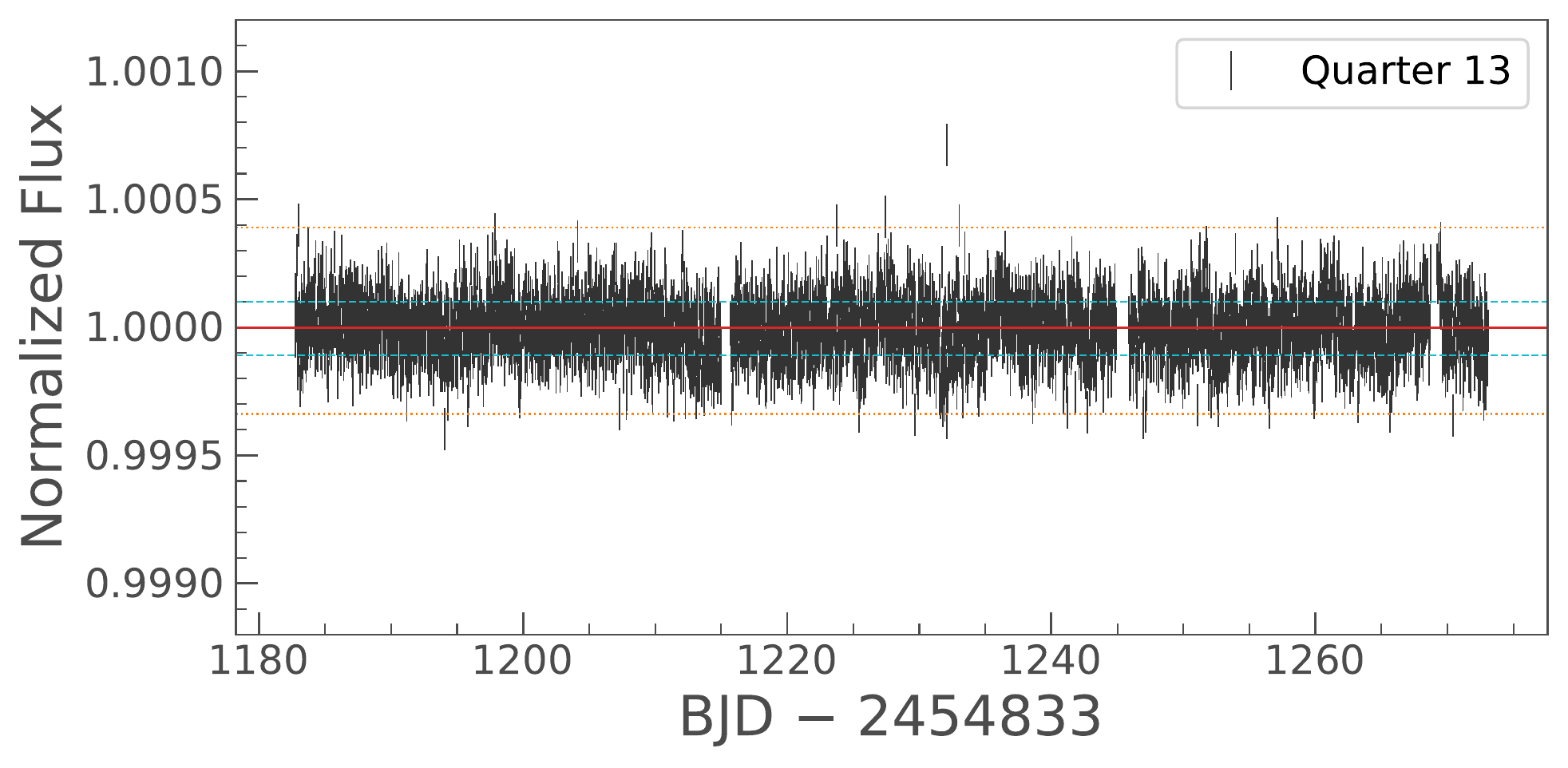} &
      \includegraphics[width=5.3cm]{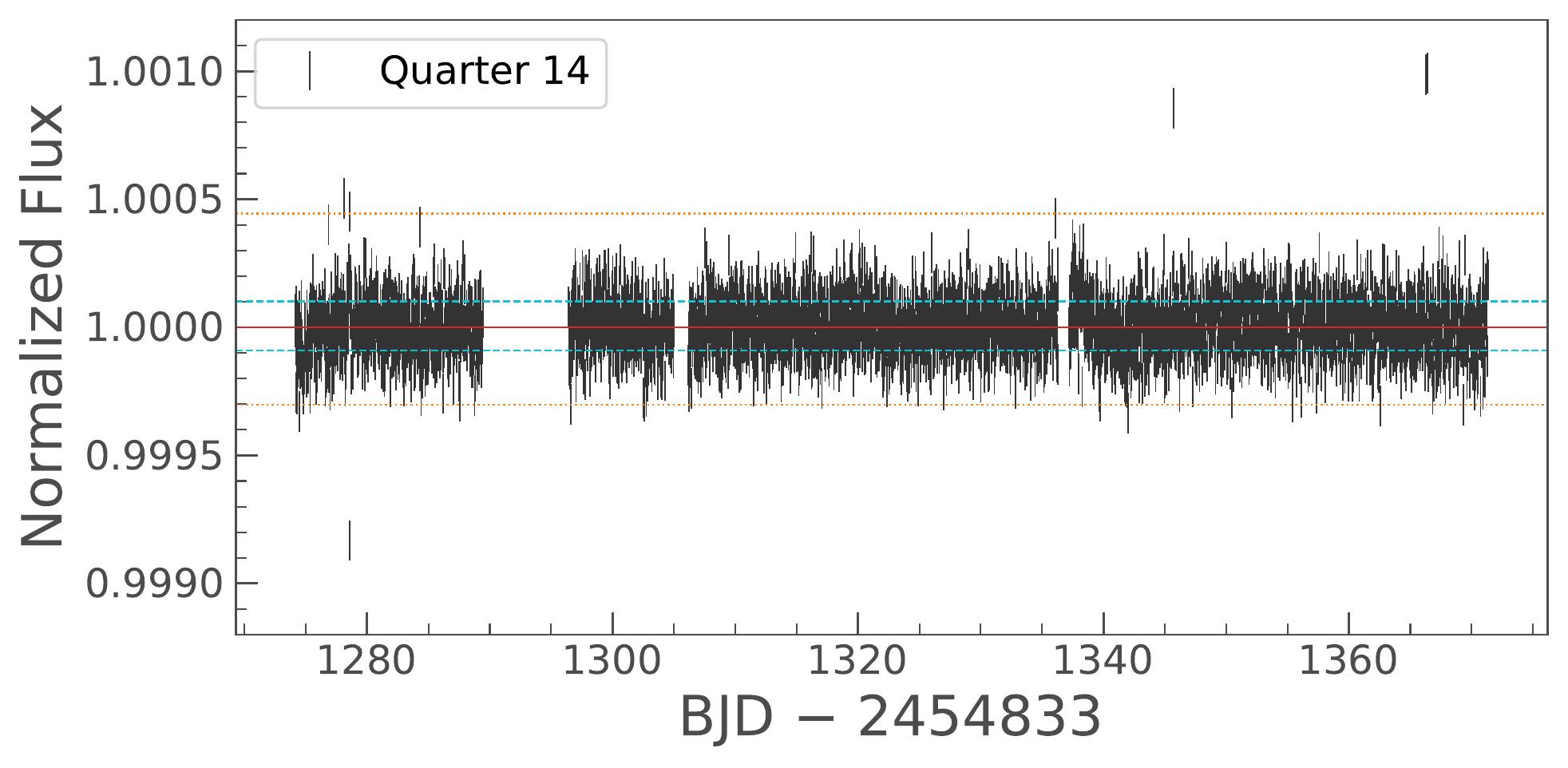} \\
      \includegraphics[width=5.3cm]{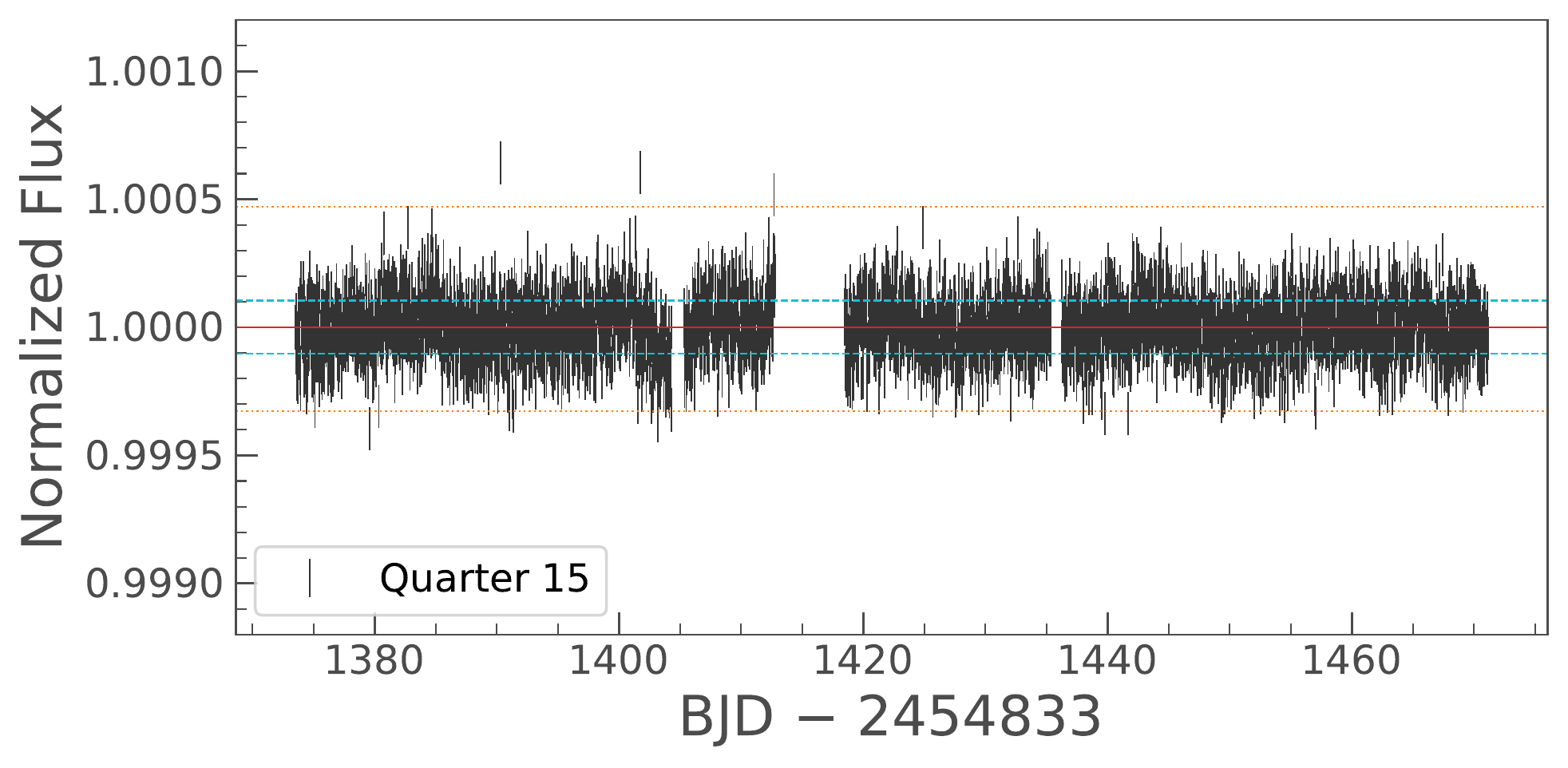} &
      \includegraphics[width=5.3cm]{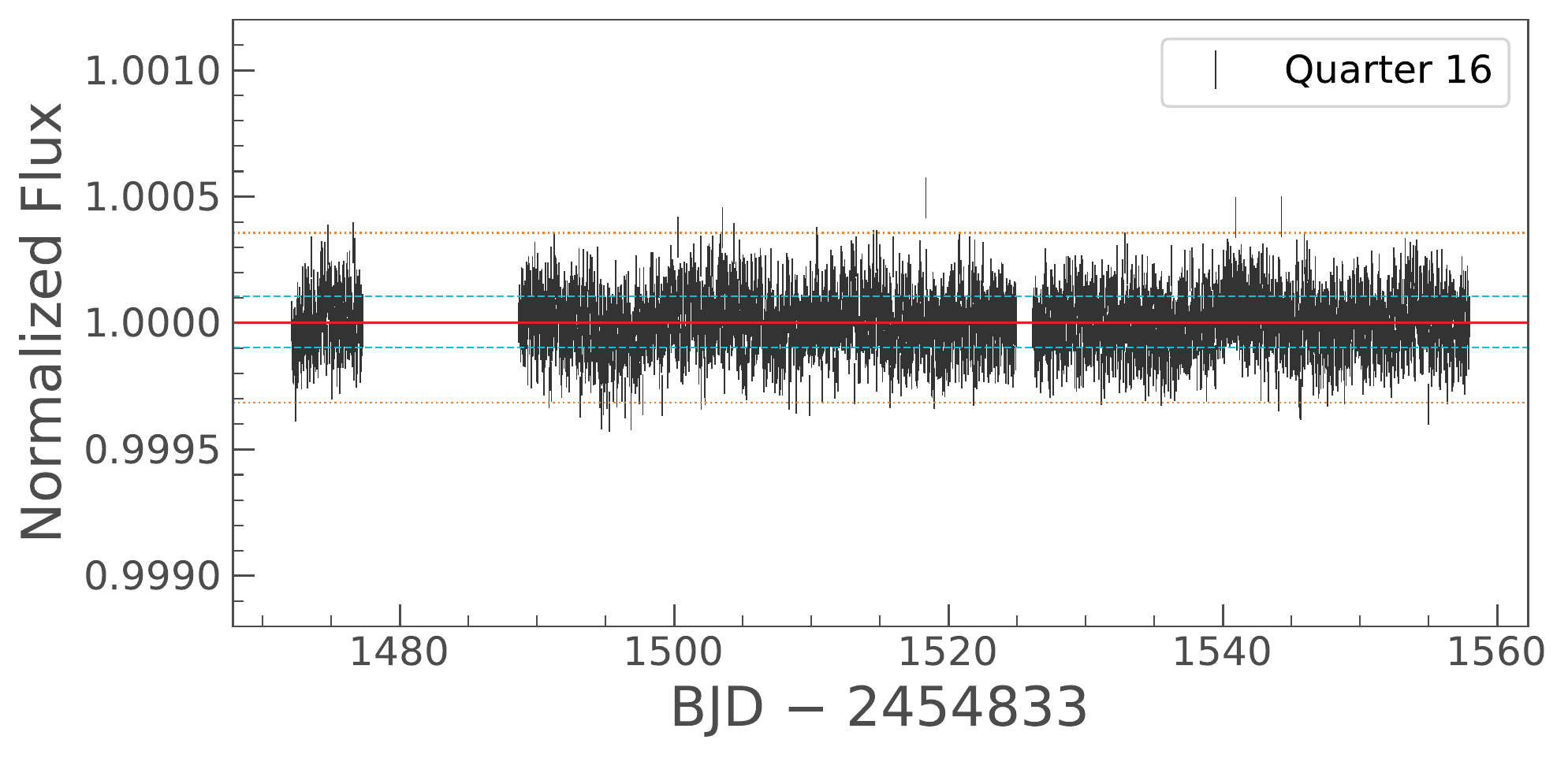} &
      \includegraphics[width=5.3cm]{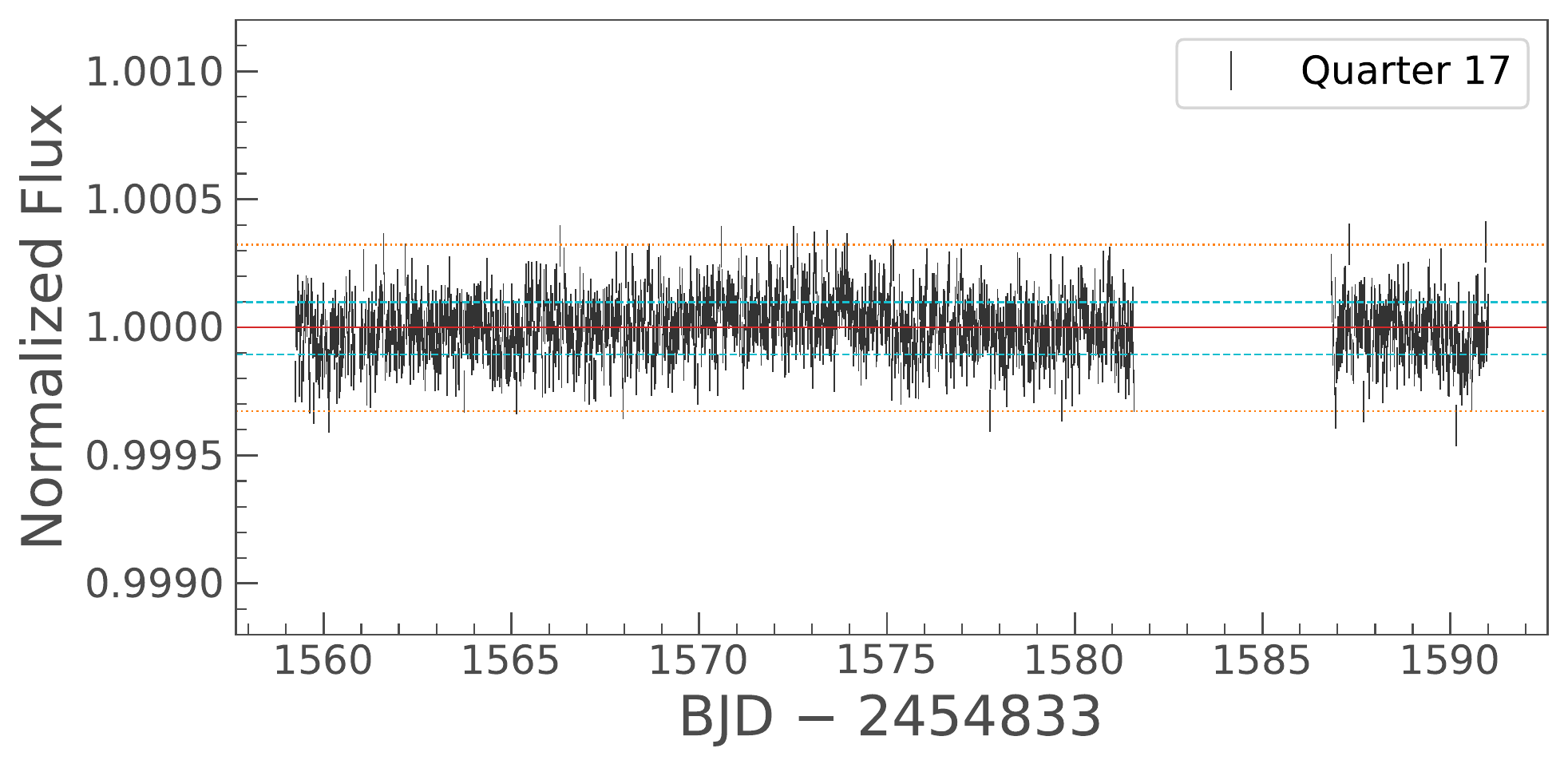} 
    \end{tabular}
  \end{center}
  \caption{Long-cadence \Kepler\ PDCSAP light curves of \kic\ from the entire \Kepler\ primary mission. The red lines denote the median of each light curve, and the blue and orange lines are plotted at percentiles that approximate the $\pm1\sigma$ and $\pm3\sigma$ fluxes, respectively. The single occultation that led to the original validation of \kicb\ is clearly visible in Quarter 4. There are no statistically appreciable counterparts to this event (i.e., transits or eclipses).}
  \label{fig:kepler}
\end{figure*}

%%%%%%%%%%%%%%%%%%%%%%%%%%%%%%%%%%%%%%%%%%%%%%%%%%%%%%%%%%%%%%%%%%%%%%%%%%

\section{Archival Data}\label{sec:archive}

The combination of high-precision photometry and AO imaging led to the initial validation of \kicb\ \citep{Wang2015}. Here, we present these archival data because they provide critical context to the reevaluation of the properties of the \kic\ system.  

\subsection{\Kepler\ Photometry}\label{sec:kepler}

The \Kepler\ spacecraft observed \kic\ in each quarter of the primary \Kepler\ mission (Figure \ref{fig:kepler}). We acquire all \Kepler\ photometry of \kic\ from the Milkuski Archive for Space Telescopes (MAST) using the \texttt{Lightkurve} package \citep{Lightkurve2018}. We use the pre-search data conditioning photometry (PDCSAP) from Data Release 25. This release, which uses version 9.3 of the Science Operations Center pipeline, benefits from an accurate determination of the stellar scene and crowding metrics that mitigates errors from so-called ``phantom stars'' \citep{Dalba2017a} and that leads to higher photometric precision in general \citep{Twicken2016}. We remove long-term variability from each quarter's PDCSAP light curve using a Savitzky--Golay filter \citep{Savitzky1964}. Figure \ref{fig:kepler} shows the resulting light curves of \kic. Each panel presents a single quarter, with horizontal lines denoting the following percentiles of the photometry: \{0.1, 15.9, 50, 84.1, 99.9\}. Visual inspection readily identifies the single occultation\footnote{We hereafter refer to the dimming event in Quarter 4 of the \Kepler\ light curves as an ``occultation'' event. As we will explain, it is not known whether this event is the result of an exoplanet transit or a binary star eclipse.} event in Quarter 4. Besides this occultation, there are no other statistically significant dimming events (i.e., transits or eclipses) in the light curves. 

\begin{figure}
    \centering
    \includegraphics[width=0.9\columnwidth]{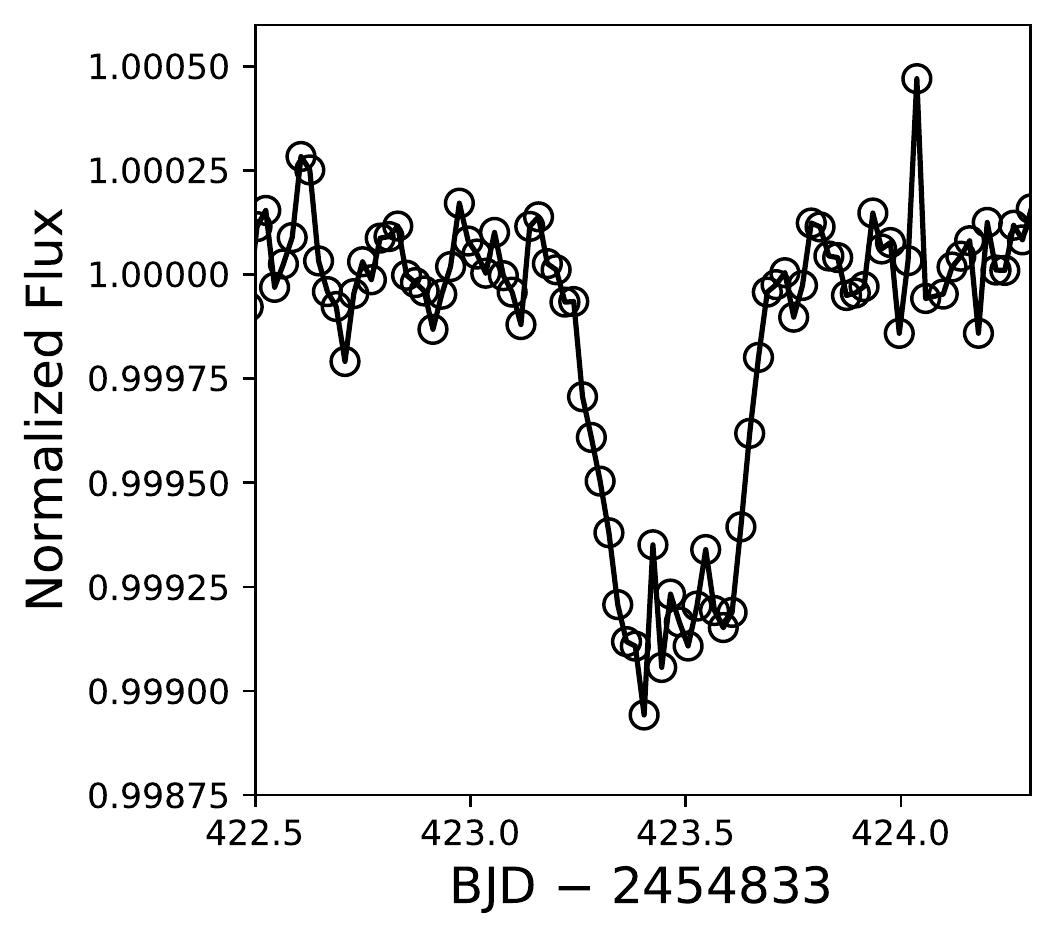}
    \caption{Single occultation event for \kic\ observed by \Kepler\ during Quarter 4 of the primary mission. The depth is consistent with a planetary transit, but the shape and duration indicate that it could be a grazing event by a larger body.}
    \label{fig:transit}
\end{figure}

Figure \ref{fig:transit} provides a detailed look at the single occultation from Quarter 4. The occultation depth ($\sim$0.1\%) is sufficiently small to suggest that the occulter is planetary in size, but the long ingress and egress durations relative to the occultation duration (i.e., second to third contact) suggest that it was possibly grazing. Indeed, a recent effort to characterize this single occultation inferred an impact parameter of 0.94$^{+0.01}_{-0.02}$ \citep{Kawahara2019}. This leads to a degeneracy between parameters that prevents the accurate estimation of the size of the occulting object.   

The duration of the single occultation event in Quarter 4 is approximately 12.3~hr. This is sufficiently long to suggest that the occulter has a long orbital period relative to the majority of known transiting exoplanets, especially considering that the occultation is likely grazing. Alternatively, the large (albeit imprecise) radius of \kic\ (see Table \ref{tab:prop}) causes a longer-duration occultation than a smaller star for a planet with a fixed orbital period. Based on analysis of the single occultation, \citet{Wang2015} estimated that the orbital period of \kicb\ is $1320.10^{+12401.80}_{-152.50}$~days. Similarly, \citet{Kawahara2019} estimated that the orbital period is $1600^{+1100}_{-400}$~days.     

\begin{deluxetable*}{lcc}
\tablecaption{Properties of the \kic\ System \label{tab:prop}}
\tablehead{
  \colhead{Property} & 
  \multicolumn{2}{c}{Value}}
\startdata
Alias                    & \multicolumn{2}{c}{Kepler-456 (b)} \\
R.A. (J2000)               & \multicolumn{2}{c}{19h 15m 57.979s} \\
Decl. (J2000)              & \multicolumn{2}{c}{+41d 13m 22.909s} \\
\Kepler\ magnitude       & \multicolumn{2}{c}{12.713} \\
Spectral type            & \multicolumn{2}{c}{F5V} \\
Distance (pc)            & \multicolumn{2}{c}{$743\pm12$} 
\smallskip \\            & \textbf{HIRES spectrum}  & \textbf{\citet{Kawahara2019}} \smallskip\\
$T_{\rm eff}$ (K)        & $5930\pm110$   & $5993^{+99}_{-88}$ \\
$R_{\star}$ ($R_{\sun}$) & $1.34\pm0.18$  & $1.81^{+0.03}_{-0.04}$ \\
$[$Fe/H$]$ (dex)         & $-0.09\pm0.09$ & $-0.05\pm0.09$ \\
$M_{\star}$ ($M_{\sun}$) & $1.16\pm0.04$  & $1.20^{+0.05}_{-0.03}$  \\
log $g$ (cgs)            & $4.10\pm0.10$  & $4.00\pm0.02$ \\
$V\sin{i}$ (km s$^{-1}$) & $3.6\pm1.0$    & \nodata \\
Age (Gyr)                & \nodata        & $4.5^{+1.0}_{-0.4}$ \\
\enddata
\tablenotetext{}{The spectral type was acquired from the SIMBAD Astronomical Database, accessed 2020 April 3. The distance was adopted from \citet{BailerJones2018}. We adopt the spectroscopic parameters of \citet{Kawahara2019} for all properties except $V\sin{i}$ (see Section \ref{sec:specprop} for an explanation).}
\end{deluxetable*}

\subsubsection{A Lower Limit on Orbital Period}\label{sec:Plim}

We do not attempt to make a new orbital period measurement from the Quarter 4 occultation event. Instead, we take advantage of the substantial baseline of the \Kepler\ photometry along with the clear nondetection of additional events to estimate a lower limit on this orbital period that does not rely on the event duration. In the simplest case, assuming that an additional occultation event could not have occurred at any time during the primary \Kepler\ mission (such as in a data gap), the shortest possible orbital period would be the time between the end of Quarter 17 and the occultation event in Quarter 4: 1168~days. 

However, we also take a more conservative approach to estimate this lower limit. For orbital periods between 1 and 1168~days, incremented by 1~day, we inject identical occultation events into the \Kepler\ light curves. The conjunction time for each period is fixed to the observed occultation timing. If less than half of an occultation occurs during a data gap, we expect that event to have been detected since the photometric precision in each quarter's light curve is well below the occultation depth. Following this procedure, candidate orbital periods will be those that only yield one detection (i.e., the actual occultation in Quarter 4). We find 56 candidate orbital periods below 1168~days, the shortest of which is 609~days. In later analyses (Section \ref{sec:rj_planet}), we will treat this value as the shortest possible orbital period of the object that caused the Quarter 4 occultation.

\subsection{NIRC2 Adaptive Optics Imaging}\label{sec:nirc2}

\begin{figure}
    \centering
    \includegraphics[width=\columnwidth]{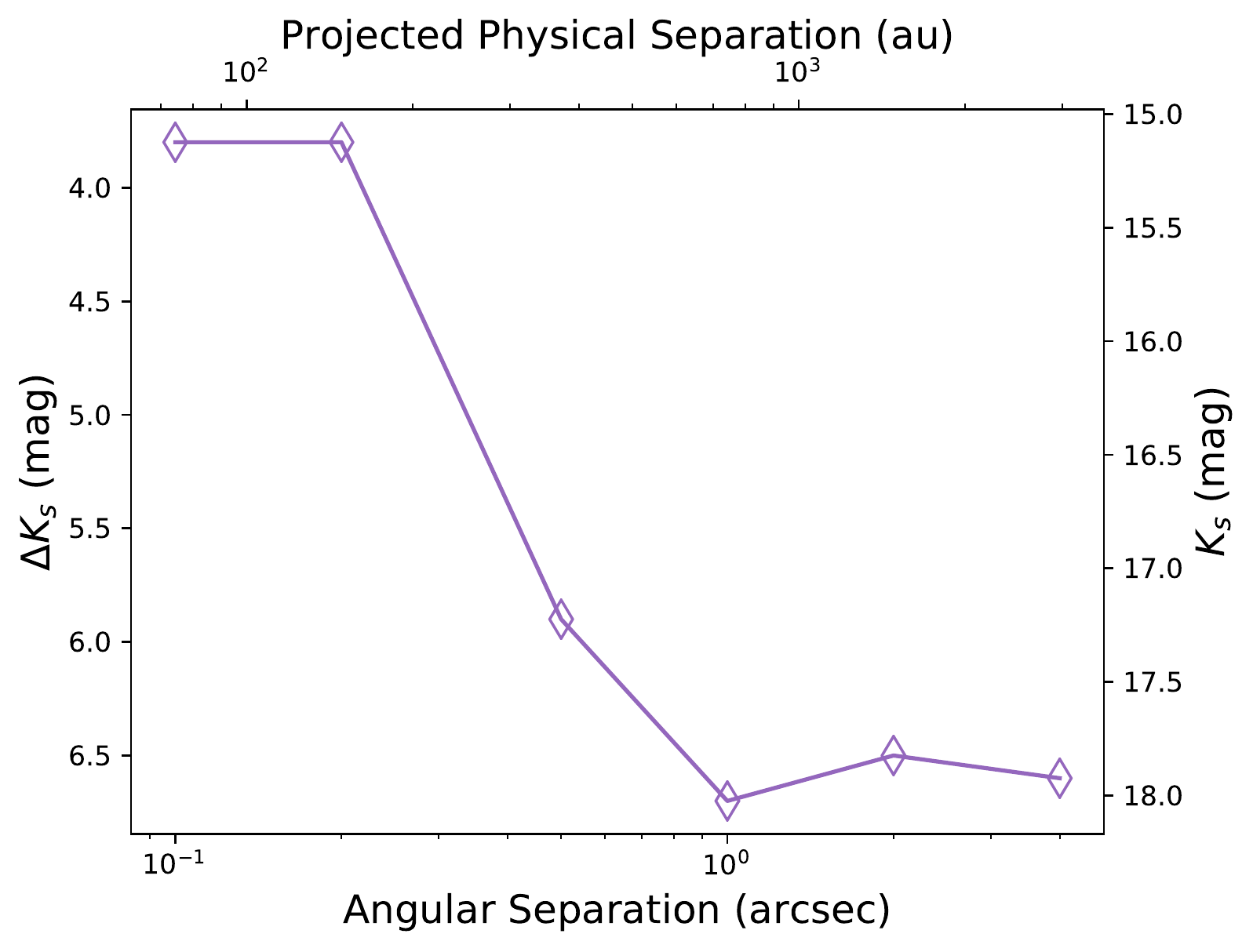}
    \caption{Limits on the magnitude of undetected stellar companions to \kic\ based on NIRC2 AO imaging \citep{Wang2015}. The detection limits are five times the standard deviation of the flux above the median in concentric annuli at the angular separations shown. The projected physical separation is simply the product of the angular separation and known distance to \kic.}
    \label{fig:nirc2}
\end{figure}

AO imaging of \kic\ was conducted using the NIRC2 instrument \citep{Wizinowich2000} on the Keck II telescope \citep{Wang2015}. Images were taken in the $K_s$ band to search for stellar companions that may have contaminated the flux in the \Kepler\ apertures or that may indicate that the occultation signal is a false-positive. No stellar companions were identified for \kic. Instead, limits were placed on the magnitude difference between a potential undetected stellar companion and \kic\ (Figure \ref{fig:nirc2}). The detection limits are five times the standard deviation of the flux above the median in concentric annuli at the angular separations shown. 

The non-detection of additional stellar components in the \kic\ AO images contributed to the confidence that the observed occultation was not some sort of false-positive scenario. However, this observation does not clearly distinguish between the scenarios of a planet transit or a grazing eclipse of a stellar companion. The distance between \kic\ and the solar system (Table \ref{tab:prop}) limits the inner edge of the AO detection limit to $\sim70$~au. Stellar or planetary companions could certainly orbit the host star well within this separation and could have the proper inclination to cause either a transit or an eclipse. Furthermore, there is also a small possibility that the AO observations occurred near the time of inferior or superior conjunction of the stellar companion.

%%%%%%%%%%%%%%%%%%%%%%%%%%%%%%%%%%%%%%%%%%%%%%%%%%%%%%%%%%%%%%%%%%%%%%%%%%

\section{Spectroscopic Observations}\label{sec:spec}

Despite the apparent grazing nature of the occultation event, the degeneracy between planetary radius and impact parameter, the weak constraint on orbital period, and the limited AO imaging, the observations of \kicb\ were sufficient to validate its planetary nature at a confidence of 99.8\% \citep{Wang2015}. In the following sections, we describe efforts to extend the analysis of \kic\ to include spectroscopic characterization and Doppler monitoring with the goal of measuring the mass and determining the planetary nature of \kicb.

\subsection{Spectroscopic Properties}\label{sec:specprop}

We began this process by acquiring an iodine-free spectrum of \kic\ with HIRES \citep{Vogt1994} on the Keck I telescope. The signal-to-noise ratio (S/N) of this spectrum is $\sim$40, which is sufficient to rule out the presence of a second set of spectral lines and to conduct a basic spectroscopic analysis of \kic. We use \texttt{SpecMatch-Emp}, an empirical spectrum matching technique \citep{Yee2017},\footnote{\url{https://specmatch-emp.readthedocs.io/en/latest/}} to infer the stellar radius ($R_{\star}$), stellar effective temperature ($T_{\rm eff}$), and iron abundance ($\rm [Fe/H]$) of \kic. \texttt{SpecMatch-Emp} does not compute stellar surface gravity ($\log{g}$), mass ($M_{\star}$), or projected rotational velocity ($V\sin{i}$). Instead, we employ the \texttt{SpecMatch} technique \citep{Petigura2015,Petigura2017} to calculate these properties. \texttt{SpecMatch} interpolates a grid of model stellar spectra and is especially suited to low-S/N, iodine-free spectra acquired with HIRES. The resulting spectroscopic properties of \kic\ are listed in Table \ref{tab:prop}. 

The properties of \kic\ place it in a region of parameter space near the edge of \texttt{SpecMatch}'s applicability. Therefore, we also list the spectroscopic properties of \kic\ from \citet{Kawahara2019} in Table \ref{tab:prop}. \citet{Kawahara2019} processed a spectrum acquired with the Large Sky Area Multi-Object Fiber Spectroscopic Telescope \citep{Cui2012,Luo2015} using \texttt{SpecMatch-Emp} and isochrone modeling. This characterization included parallax measurements from the \Gaia\ mission \citep{BailerJones2018,Gaia2018}. Since the applicability of \texttt{SpecMatch} to \kic\ is questionable, we hereafter adopt all stellar properties of \kic\ from \citet{Kawahara2019} except for $V\sin{i}$, which was not reported from that analysis.

%%%%%%%%%%%%%%%%%%%%%%%%%%%%%%%%%%%%%%%%%%%%%%%%%%%%%%%%%%%%%%%%%%%%%%%%%%

\section{Doppler Monitoring and the Matched-template Technique}\label{sec:match_temp}

To explore possible false-positive scenarios for the \kicb\ occultation event, we conducted a Doppler monitoring campaign with HIRES on the Keck I telescope. We acquired six high-resolution ($R\approx60,000$) spectra of \kic. A heated iodine cell in the light path in front of the slit enables precise wavelength calibration of each RV measurement. The observed spectrum, which is the combination of the stellar and the gaseous iodine spectra, is then forward modeled following standard procedures of the California Planet Search \citep[e.g.,][]{Howard2010,Howard2016}. 

Iodine-calibrated RV measurements have a long history of producing precise velocities \citep{Marcy1992} but one of the primary limitations of the technique is the need to acquire a high-S/N, iodine-free spectrum of the star. This ``template'' observation is deconvolved from the instrumental point spread function (PSF) that is derived from bracketing observations of rapidly rotating B stars (taken with the iodine cell) to construct a ``deconvolved stellar spectral template'' (DSST). This deconvolution requires the template to be high-S/N, ideally near 200 per pixel. During the forward modeling process, the DSST is multiplied by an ideal iodine absorption spectrum as measured with a Fourier transform spectrograph and then convolved with the PSF derived for each observation. Since HIRES is slit-fed, the PSF is highly dependent on the precise illumination pattern of the slit, which is affected by changes in seeing and guiding. The variable PSF requires a complex model with 12 free parameters that is capable of modeling tiny asymmetries in the PSF. The large number of free parameters combined with noise amplification in the deconvolution process drives the need for high S/N ($\sim$200) for high RV precision ($\sim$1~m~s$^{-1}$). This makes it impractically expensive to observe faint targets like \kic, which would require an hour-long exposure to reach an S/N of only $\sim$100.

\citet{Fulton2015b} developed a technique to utilize model stellar spectra in place of the DSST and showed that RV precision of $\sim$5--15~m~s$^{-1}$ precision could be achieved without observing a template of the target star. In addition, the noise-free (model) template spectrum allows for much lower S/N for each RV observation. However, this technique is limited to the relatively narrow regime of FGK main-sequence dwarf stars, where the \citet{Coelho2005} models are sufficiently accurate. 

\begin{figure}
    \centering
    \includegraphics[width=\columnwidth]{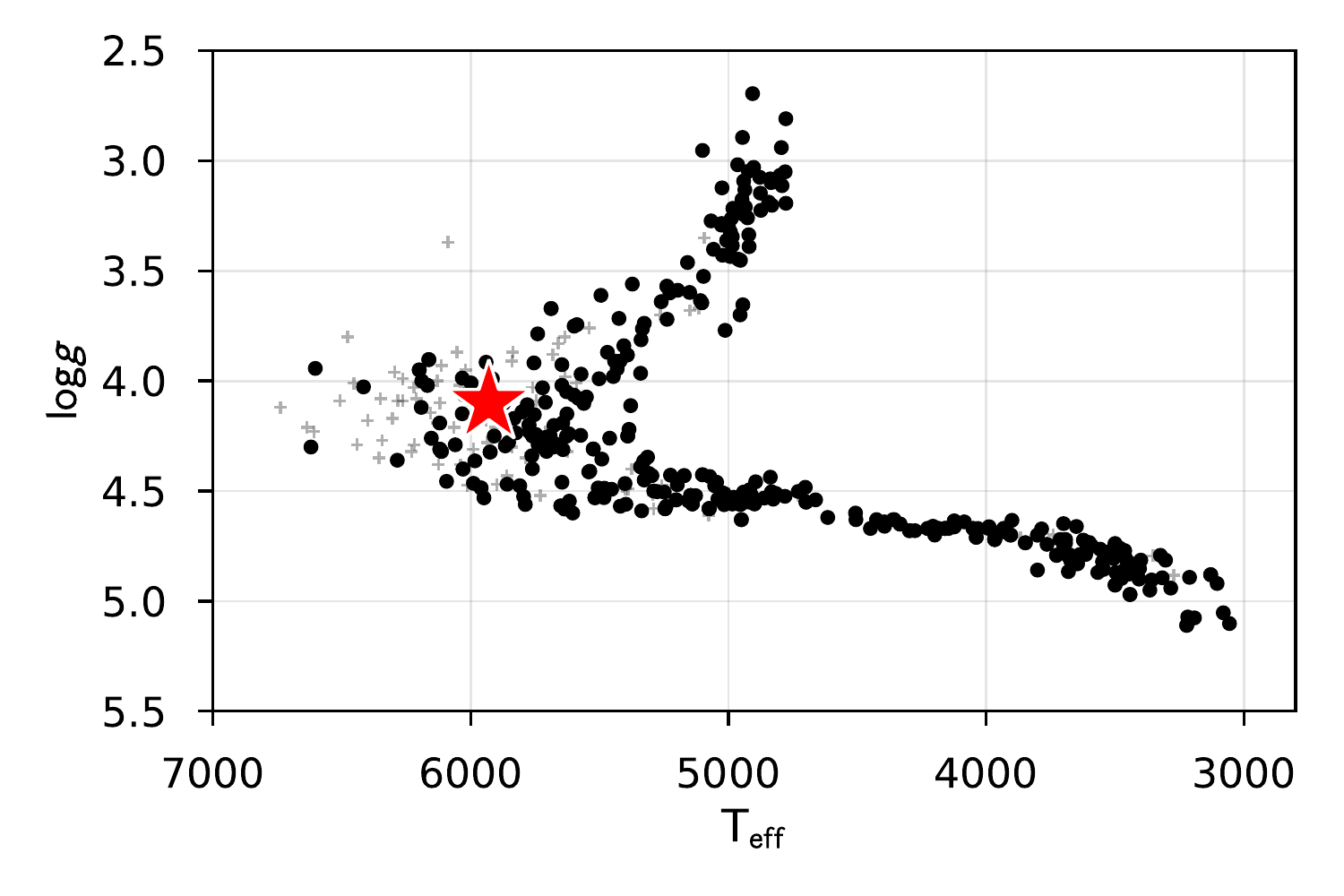}
    \caption{Spectroscopic H-R diagram of the \texttt{SpecMatch-Emp} library. \kic\ is represented by the red star. Stars with DSSTs are plotted as black circles, and library stars without DSSTs are plotted as gray plus signs. Our DSSTs span F to M dwarfs, as well as the majority of the subgiant branch.}
    \label{fig:hr}
\end{figure}

\citet{Yee2017} utilized decades of archival HIRES data to construct a library of high-quality template spectra and developed a technique to match these spectra to an observed, low-S/N, iodine-free observation to extract precise stellar parameters (\texttt{SpecMatch-Emp}). In this work, we use this library and spectral matching technique to identify a spectrum in the California Planet Search archive that can be substituted for the DSST. For any star without a DSST, we identify the member of the library with the most similar spectrum as quantified by the absolute $\chi^2$ statistic. Some of the stars in the \citet{Yee2017} template library were not observed with bracketing B stars and therefore cannot be used to derive DSSTs. Figure \ref{fig:hr} plots the \texttt{SpecMatch-Emp} library on a spectroscopic Hertzprung-Russell (HR) diagram and shows the distribution of library stars with and without DSSTs. Fortunately, 323 of the 404 library stars were observed appropriately and have associated DSSTs in the HIRES database. The stars have effective temperatures spanning from $\sim$6500~K to $\sim$3000~K and $\log{g}$ from $\sim$5 to 2.5~dex, making this technique possible for the majority of main-sequence and subgiant stars amenable to precision RVs.

\subsection{Validation of the Matched-template Technique}\label{sec:validate}

We validate the matched-template technique by applying it to the HIRES template library. Since each star in the library has its own template, its RVs can be used as a ``ground truth'' in the comparison with RVs produced via the matched-template technique. This comparison can also be conducted as a function of stellar properties to additionally explore the sensitivity and limitations of the technique. 

\begin{figure*}
    \centering
    \includegraphics[width=0.9\textwidth]{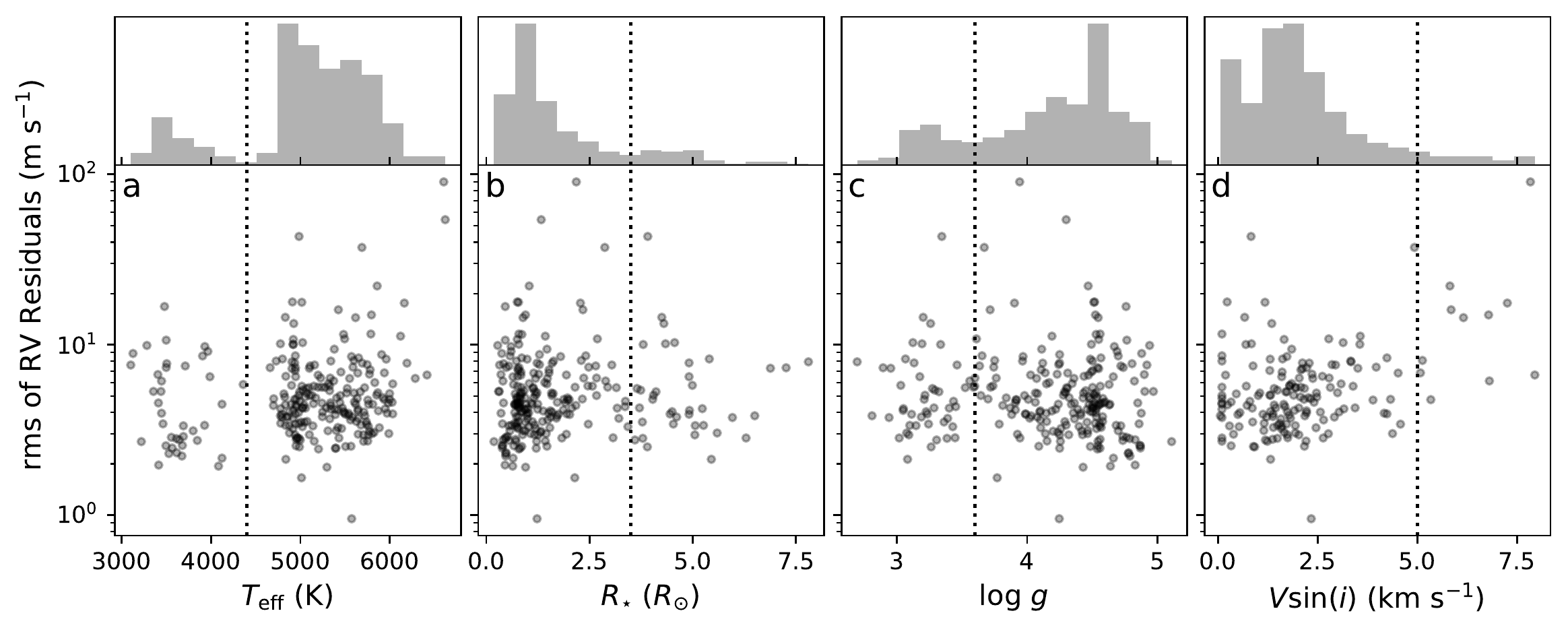}
    \caption{Stellar properties of the 247 stars in the HIRES DSST library on which the matched-template technique is tested and the corresponding rms of the RV residuals (Equation \ref{eq:rms}). These panels show that the matched-template technique can be applied to a wide variety of stars. However, we identify groups of stars of various properties that have slightly different distributions of rms values (shown in Figure \ref{fig:rms_dist}). We separate the groups with dotted lines, the values of which are given in Table \ref{tab:rms}.}
    \label{fig:rms_relative}
\end{figure*}

We begin by identifying the subset of the 323 stars in the DSST library that are appropriate for this test. We require that stars have at least three iodine-in spectral observations. This criterion removes nine stars. We also require that each star can be suitably matched with another in the library. This criterion removes 67 stars, most of which reside in relatively isolated regions of parameter space. This leaves 247 stars for the test of the matched-template technique. The range of stellar properties of these stars is displayed in Figure \ref{fig:rms_relative}. 

For each of the 247 stars, we select up to the 100 spectra with the highest S/N. We process each of these spectra twice: once using the template of the star itself, and once using the best-match star's template. The best-match star is chosen to be the one with the most similar spectrum, as quantified by the absolute $\chi^2$ statistic. Hereafter, we will use the subscripts ``self'' and ``match'' to refer to RV data products produced using a star's own template and its best-match star's template, respectively. For each star, we calculate the rms of the residuals between the two RV time series as
\begin{equation}\label{eq:rms}
    {\rm rms} = \sqrt{\frac{\sum_{i=1}^n [v_{\rm r,match}(t_i) - v_{\rm r,self}(t_i)]^2}{n}}  
\end{equation}

\noindent where $v_{\rm r}$ is the stellar RV, $t$ is time, and $n$ is the total number of RV observations for the $i$th star. The median number of RVs for each star used in the calculation of the rms is 26, and the mean number of RVs is 37.

\begin{figure}
    \centering
    \includegraphics[width=\columnwidth]{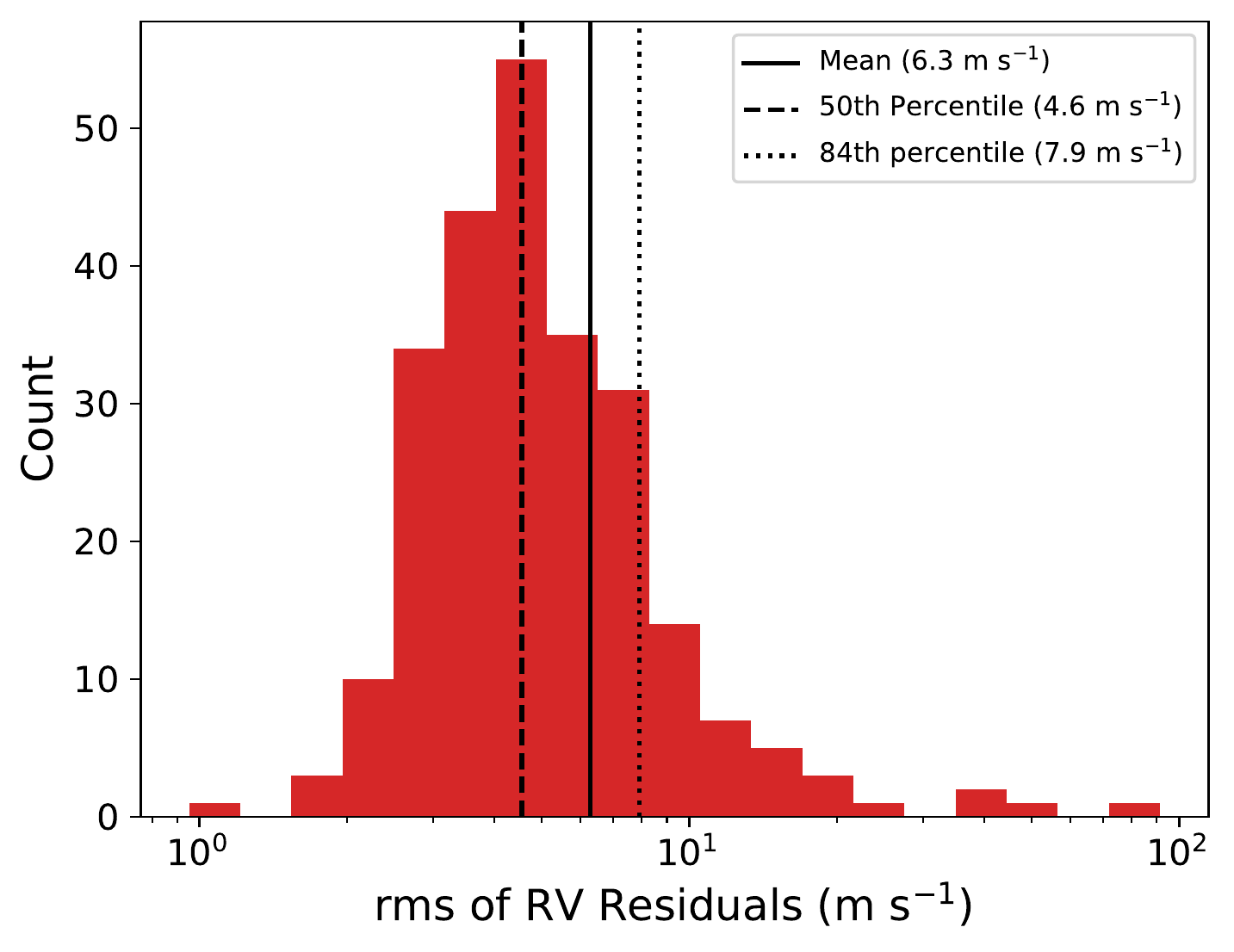}
    \caption{Distribution of rms values of the RV residuals (Equation \ref{eq:rms}) for the 247 stars tested with the matched-template method. The various statistics provided for this distribution serve as a general assessment of the precision of the matched-template method.}
    \label{fig:resid_rms}
\end{figure}

Figure \ref{fig:resid_rms} shows the distribution of rms values of the RV residuals for all 247 stars. The logarithms of the rms values form a somewhat normal distribution that is skewed toward higher values. The median of this distribution is 4.6~m~s$^{-1}$, which serves as a first-order estimate of the noise floor of the matched-template method. On average, the rms of the residuals between the $v_{\rm r,match}$ and $v_{\rm r,self}$ time series is larger than the median uncertainty on $v_{\rm r,self}$. We therefore cannot neglect the error that is introduced by the matched-template analysis.

We consider how the method affects the uncertainty of the derived RV measurements as a function of stellar properties. In Figure \ref{fig:rms_relative}, the rms of the RV residuals is plotted against $T_{\rm eff}$, $R_{\star}$, $\log{g}$, and $V\sin{i}$. The range of the stellar properties spanned in this figure shows that the matched-template method can be applied to many different types of stars. However, different types of stars produce slightly different distributions of rms values (as separated by dotted lines in Figure \ref{fig:rms_relative}). 

\begin{figure}
    \centering
    \includegraphics[width=\columnwidth]{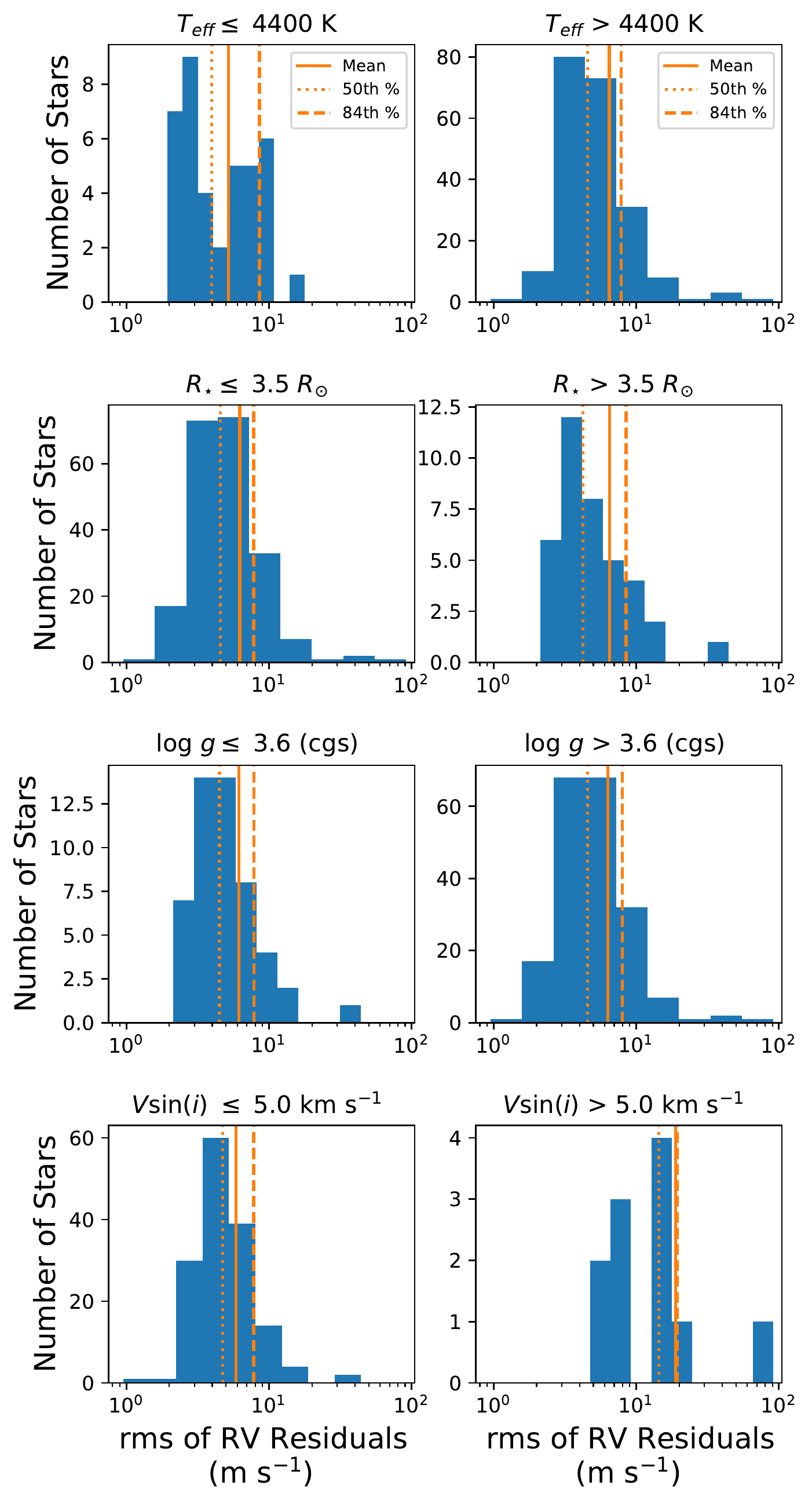}
    \caption{Distributions of rms of RV residuals for each group in each of four stellar parameters. The divisions listed here are represented as the dotted lines in Figure \ref{fig:rms_relative}. In general, these distributions are unimodal with tails to longer values. The only distributions that are not also have relatively few stars. The values of the orange vertical lines shown here are listed in Table \ref{tab:rms} and are useful for determining how to inflate the internal error for RVs produced with the matched-template method.}
    \label{fig:rms_dist}
\end{figure}

For each of the four stellar properties we consider, we divide the stars into two groups. The locations of these divisions are chosen manually and are meant to separate groups of stars with similar properties and also different distributions of rms values. The values of the divisions in stellar properties are listed in Table \ref{tab:rms}. The resulting distributions of the rms of the RV residuals are shown in Figure \ref{fig:rms_dist}. In general, these distributions are unimodal but have tails to larger values. The only distributions that are not unimodal (i.e., $T_{\rm eff} \le 4400$~K and $V\sin{i}>5.0$~km~s$^{-1}$) contain substantially fewer stars than the others. This demonstrates that for cool stars, or those with relatively high rotation velocity, the matched-template method will have limited accuracy owing to the limited sample of stars in the HIRES template library. For all other distributions, Figure \ref{fig:rms_dist} identifies the mean, 50th percentile (i.e., median), and 84th percentile of the rms of the RV residual values. These values are key to understanding how the matched-template method inflates the RV uncertainty. 

Table \ref{tab:rms} contains the statistics describing the distributions in Figure \ref{fig:rms_dist} and is meant to be a ``look-up table.'' For a given star---with known $T_{\rm eff}$, $R_{\star}$, $\log{g}$, and $V\sin{i}$---four estimates of the rms of the RV residuals for similar stars can be identified (one for each stellar property). The estimates can each be the mean, median, or 84th percentile, the choice of which depends on the particular application of the matched-template method and the error tolerance. Those four estimates of rms can then be combined (e.g., minimum, maximum, median, etc.) to yield a final error values to be added (in quadrature) to the RV internal errors. For example, for a solar analog star, Table \ref{tab:rms} yields median values of 4.6~m~s$^{-1}$ (for $T_{\rm eff}$), 4.6~m~s$^{-1}$ (for $R_{\star}$), 4.6~m~s$^{-1}$ (for $\log{g}$), and 4.7~m~s$^{-1}$ (for $V\sin{i}$), the average of which is 4.6~m~s$^{-1}$. Based on the needs of the analysis, more or less conservative approaches may be justified, and the values listed in Table \ref{tab:rms} allow for those. 

\begin{deluxetable*}{lccc}
\tablecaption{Error Incurred by the Matched-template Technique as a Function of Stellar Properties \label{tab:rms}}
\tablehead{
  \colhead{} & 
  \multicolumn{3}{c}{\underline{Statistics Describing rms of RV Residual Distribution (m s$^{-1}$)}}\\
  \colhead{Stellar Properties} &
  \colhead{~~~~~~~~~~Mean~~~~~~~~~~} &
  \colhead{50th Percentile} &
  \colhead{84th Percentile}}
\startdata
Effective temperature (K)         &&& \\
~~~~$T_{\rm eff} \le 4400$         & 5.2 & 4.0 & 8.6 \\
~~~~$T_{\rm eff} > 4400$           & 6.5 & 4.6 & 7.8 \\
Radius ($R_{\sun}$)               &&& \\
~~~~$R_{\star} \le 3.5$            & 6.2 & 4.6 & 7.8 \\
~~~~$R_{\star} > 3.5$              & 6.5 & 4.2 & 8.5 \\
Surface gravity (cgs)             &&& \\
~~~~log $g \le 3.6$                & 6.1 & 4.5 & 7.9 \\
~~~~log $g > 3.6$                  & 6.3 & 4.6 & 8.0 \\
Rotational velocity (km s$^{-1}$) &&& \\
~~~~$V\sin{i} \le 5.0$             & 5.9  & 4.7 & 14.4 \\
~~~~$V\sin{i} > 5.0$               & 18.9 & 7.8 & 19.5 \\
\enddata
\tablenotetext{}{When applying the matched-template method to a new star, identify the applicable rms values based on its properties. Choose a suitable method of combining these rms values (e.g., their median, maximum, etc.), and add the final value \textit{in quadrature} to the internal RV errors.}
\end{deluxetable*}

For most stars that would be subject to precise RV measurements, the matched-template method demonstrates the ability to surpass the $\sim$10~m~s$^{-1}$ noise floor of the synthetic template technique \citep{Fulton2015b} by nearly a factor of two. For faint stars for which acquiring a high-S/N template is infeasible, assuming that there is a suitable best-match star in the HIRES DSST library, the matched-template technique is preferable. Furthermore, the method is also useful for exploring the RV signals of bright stars prior to the acquisition of a template spectrum. It is typical for several iodine spectra to be acquired prior to a template in order to establish a time baseline for the star of interest. The matched-template method enables a first-order assessment of the RVs associated with those iodine-in spectra that aids in planning and has the potential to save observing time.

\subsection{Application of the Matched-template Technique to \kic}

We processed the RV observations of \kic\ from Keck-HIRES with the matched-template technique using the DSST for HD~22484. Table \ref{tab:spec_comp} lists the properties of HD~22484 inferred from \texttt{SpecMatch} using its high-S/N iodine-free template spectrum. All properties between the two stars are consistent to within the $1\sigma$ uncertainties except stellar radius. 

\begin{deluxetable}{lc}
\tablecaption{Spectroscopic Properties of HD~22484, the Best Match to \kic \label{tab:spec_comp}}
\tablehead{
  \colhead{Property} & 
  \colhead{HD~22484}}
\startdata
Spectral type            & F9IV-V \\
$T_{\rm eff}$ (K)        & $5964\pm100$   \\
$R_{\star}$ ($R_{\sun}$) & $1.66\pm0.04$  \\
$[$Fe/H$]$ (dex)         & $-0.01\pm0.06$  \\
$M_{\star}$ ($M_{\sun}$) & $1.13\pm0.07$    \\
log $g$ (cgs)            & $4.07\pm0.10$  \\
$V\sin{i}$ (km s$^{-1}$) & $3.29\pm1.0$ \\
\enddata
\tablenotetext{}{All parameters other than spectral type were inferred from the iodine-free template spectrum with \texttt{SpecMatch}. The spectral type was acquired from the SIMBAD Astronomical Database (accessed 2020 April 3).}
\end{deluxetable}

The Keck-HIRES RVs resulting from the matched-template analysis of \kic\ are listed in Table \ref{tab:rv} and are hereafter assigned to the symbol $v_{\rm r}$. The internal errors on $v_{\rm r}$ are $\sim$3~m~s$^{-1}$. We follow the procedure outlined in Section \ref{sec:validate} to determine the amount by which we must increase these errors to account for the matched-template method. Based on the properties of \kic\ in Table \ref{tab:prop}, we identify rms of the RV residuals values of 6.5~m~s$^{-1}$ (for $T_{\rm eff}$), 6.2~m~s$^{-1}$ (for $R_{\star}$), 6.3~m~s$^{-1}$ (for $\log{g}$), and 5.9~m~s$^{-1}$ (for $V\sin{i}$) from Table \ref{tab:rms}. We have chosen to use the mean values of the rms distributions as a conservative trade-off between the 50th and 84th percentiles. The average of these four values is 6.2~m~s$^{-1}$, which we add in quadrature to the internal RV errors. The resulting uncertainty in $v_{\rm r}$ is denoted as $\sigma_{v_{\rm r}}$ and is listed in Table \ref{tab:rv}.

\begin{deluxetable*}{cccccc}
\tablecaption{Keck-HIRES RV Measurements for \kic \label{tab:rv}}
\tablehead{
  \colhead{Time} & 
  \colhead{Telluric RV\tablenotemark{a}  $v_{\rm t}$} &
  \colhead{$\sigma_{v_{\rm t}}$} &
  \colhead{Precise RV  $v_{\rm r}$} &
  \colhead{$\sigma_{v_{\rm r}}$\tablenotemark{b}} &
  \colhead{$S_{\rm HK}$} \\
  \colhead{(BJD$_{\rm TDB}$)} &
  \colhead{(km s$^{-1}$)} &
  \colhead{(km s$^{-1}$)} &
  \colhead{(m s$^{-1}$)} &
  \colhead{(m s$^{-1}$)} &
  \colhead{}
  }
\startdata
2458386.74225 & $-22.39$ & 0.10 & $1463.4$  & 6.8 & $0.130\pm0.001$ \\
2458393.88630 & $-22.47$ & 0.10 & $1400.1$  & 7.0 & $0.133\pm0.001$ \\
2458622.98313 & $-24.09$ & 0.10 & $-178.6$  & 6.8 & $0.136\pm0.001$ \\
2458659.08536 & $-24.62$ & 0.10 & $-447.5$  & 6.8 & $0.130\pm0.001$ \\
2458723.98153 & $-25.13$ & 0.10 & $-910.5$  & 6.7 & $0.124\pm0.001$ \\
2458780.81064 & $-25.46$ & 0.10 & $-1324.8$ & 6.7 & $0.126\pm0.001$ \\
\enddata
\tablenotetext{a}{The telluric-calibrated, absolute RVs were calculated using the methodology of \citet{Chubak2012}.}
\tablenotetext{b}{The values of $\sigma_{v_{\rm r}}$ include the 6.2~m~s$^{-1}$ error from the matched-template technique (Section \ref{sec:validate}).}
\end{deluxetable*} 

A corresponding $S_{\rm HK}$ activity indicator is listed with each RV measurement. The HIRES spectra include the Ca II H and K lines, which enable the calculation of the $S_{\rm HK}$ activity indicators \citep{Wright2004,Isaacson2010}. We find that the Pearson correlation coefficient between the RVs and $S_{\rm HK}$ values is 0.54, and the corresponding two-tailed $p$-value is 0.27. 

We also list the low-precision, telluric-calibrated RVs for \kic\ in Table \ref{tab:rv}. These measurements and their corresponding uncertainties are given the symbols $v_{\rm t}$ and $\sigma_{v_{\rm t}}$, respectively. These RVs were calculated following the methodology of \citet{Chubak2012}. They are absolute in that they share a common zero-point with other studies \citep{Latham2002,Nidever2002}. The telluric RVs of \kic\ are necessary for analysis conducted in Section \ref{sec:rj_eb}. 

\begin{figure}
    \centering
    \includegraphics[width=\columnwidth]{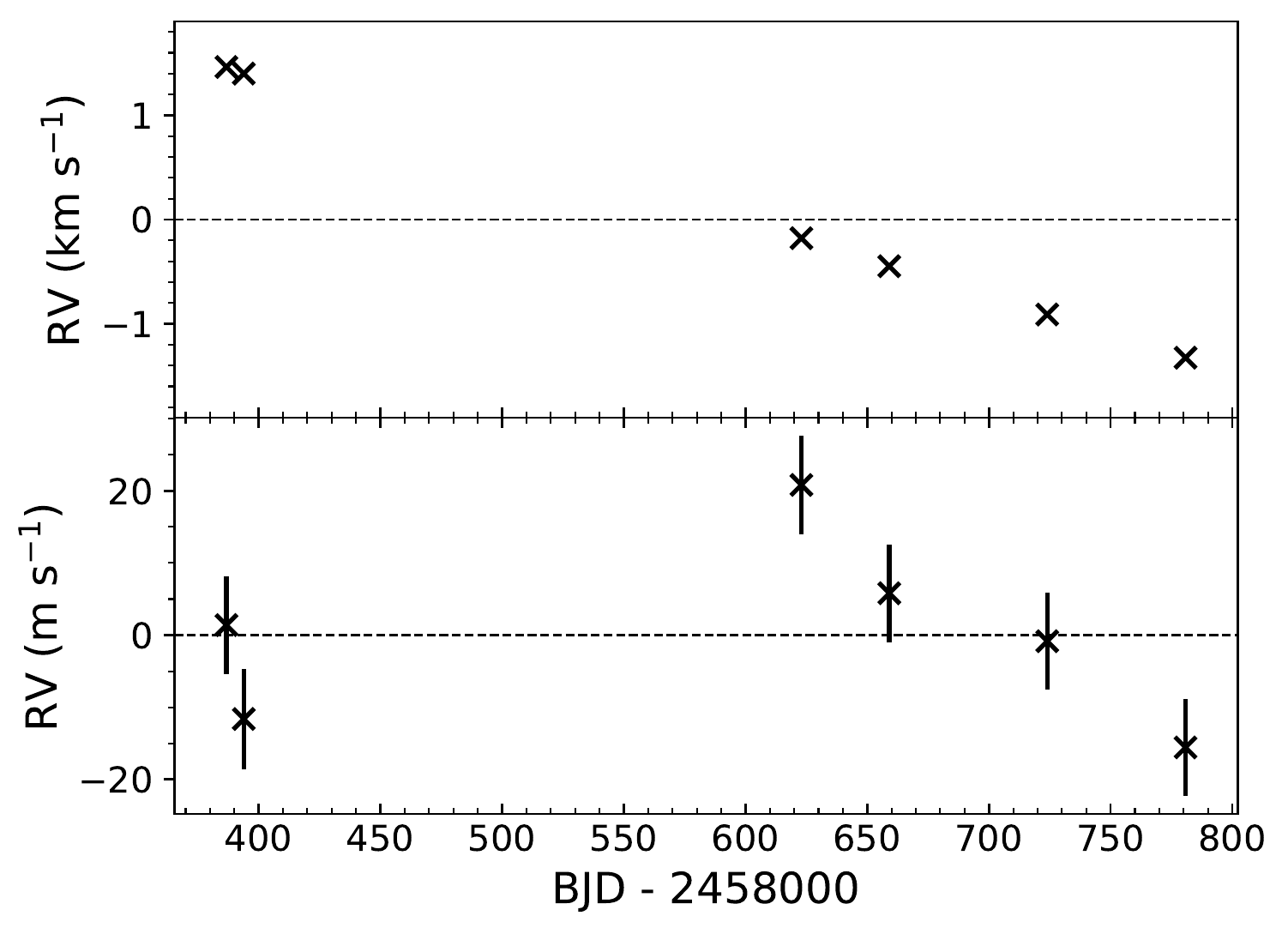}
    \caption{Top: Keck-HIRES RV measurements of \kic\ extracted using the matched-template analysis. Bottom: RV measurements after removing the trend with a linear regression. Note the difference in scale between the two panels. The errors shown here and used in this analysis include the uncertainty incurred by the matched-template technique (see Section \ref{sec:validate}).}
    \label{fig:rv}
\end{figure}

Upon first glance, the RV time series of \kic\ displays a trend on the order of several kilometers per second (Figure \ref{fig:rv}). These RV measurements are extremely linear over the 394-day baseline; the Pearson correlation coefficient between time and RV is $-$0.99994. This trend is indicative of a massive companion on a long-period orbit. When combined with the shape and duration of the occultation event in the \Kepler\ photometry, the RVs may suggest that \kicb\ is a misidentified grazing, eclipsing binary. However, a closer inspection of the RVs uncovers an additional (potentially Keplerian) signal on top of the large trend. Could there be a planet in this binary system that indeed caused the occultation event observed by \Kepler?  

%%%%%%%%%%%%%%%%%%%%%%%%%%%%%%%%%%%%%%%%%%%%%%%%%%%%%%%%%%%%%%%%%%%%%%%%%%

\section{Rejection Sampling Analysis}\label{sec:rj}

With only six RV observations, a single occultation event, and a nondetection in the AO imaging, we cannot uniquely determine the properties and architecture of the \kic\ system. However, we can explore a wide range of stellar, substellar, and planetary solutions that are consistent with these data to make useful inferences and predictions. 

We model the RV observations with the \texttt{The Joker} \citep{PriceWhelan2017}. \texttt{The Joker} is a Monte Carlo sampler that specifically models RV observations of two-body systems. This package employs a specific case of rejection sampling whereby the prior probability distribution functions (PDFs) of the model parameters are densely sampled and their likelihood is used as the rejection scalar. \texttt{The Joker} is able to sample the posterior PDF of the model parameters despite the complex nature of the likelihood function and despite sparse or low-precision data.

By construction, the choice of prior PDFs for the model parameters is critical to the rejection sampling analysis. In all cases, we employ the default priors of \texttt{The Joker} \citep{PriceWhelan2017}. For the companion orbital period ($P$), the prior PDF is uniform in the natural log of $P$ between some minimum ($P_{\rm min}$) and maximum ($P_{\rm max}$) values. The values of $P_{\rm min}$ and $P_{\rm max}$ are chosen on a case-by-case basis as described below. For the companion orbital eccentricity ($e$), the prior PDF is a beta distribution with shape parameters $s_1=0.867$ and $s_2=3.03$. This particular beta distribution is empirically motivated by observations of RV exoplanets \citep{Kipping2013a}. For the argument ($\omega$) and phase ($\phi_p$) of periastron, the prior PDFs are uniform over the domain (0, 2$\pi$). The semiamplitude ($K$) and the systemic velocity ($\gamma$) vary linearly with the RV and are treated differently than the previous four. The prior PDFs for $K$ and $\gamma$ are assumed to be broad Gaussian functions, such that they are essentially uniform over the range of interest \citep{PriceWhelan2017}. Lastly, for all cases we hold the RV jitter fixed at 0~m~s$^{-1}$. 

In the following sections, we divide our analyses based on the different signals in the RV data. First, in Section \ref{sec:rj_planet}, we use \texttt{The Joker} to remove the long-term RV trend and characterize the potential planetary signal (Figure \ref{fig:rv}, bottom panel). The plausibility of a planetary culprit of the occultation event is also considered. Then, in Section \ref{sec:rj_eb}, we use \texttt{The Joker} to characterize the long-term RV trend (Figure \ref{fig:rv}, top panel), ignoring the planetary signal. We also consider the possibility that the \Kepler\ occultation is actually the result of an eclipsing binary system. Finally, in Section \ref{sec:rj_results}, we synthesize the results of both rejection sampling analyses.

\subsection{The Potential Planetary Signal}\label{sec:rj_planet}

To investigate the potential planetary signal in the RV measurements of \kic\ (Figure \ref{fig:rv}, bottom panel), we use \texttt{The Joker} with an additional parameter for the first-order acceleration of the companion ($\dot \gamma$). We account for the trend in the data using the median value of the posterior PDF for $\dot \gamma$. In this application of \texttt{The Joker}, the prior on orbital period is bounded by $P_{\rm min} = 10$~days and $P_{\rm max} = 100,000$~days. These value are chosen by iterating over increasingly wider domains in $P$ until the shape of the posterior PDF shows no appreciable changes. We use \texttt{The Joker} to sample the prior PDFs $2^{21}$ times. Of these, 38,466 samples of the posterior PDFs survive. 

\begin{figure*}
  \begin{center}
    \begin{tabular}{cc}
      \includegraphics[width=0.9\columnwidth]{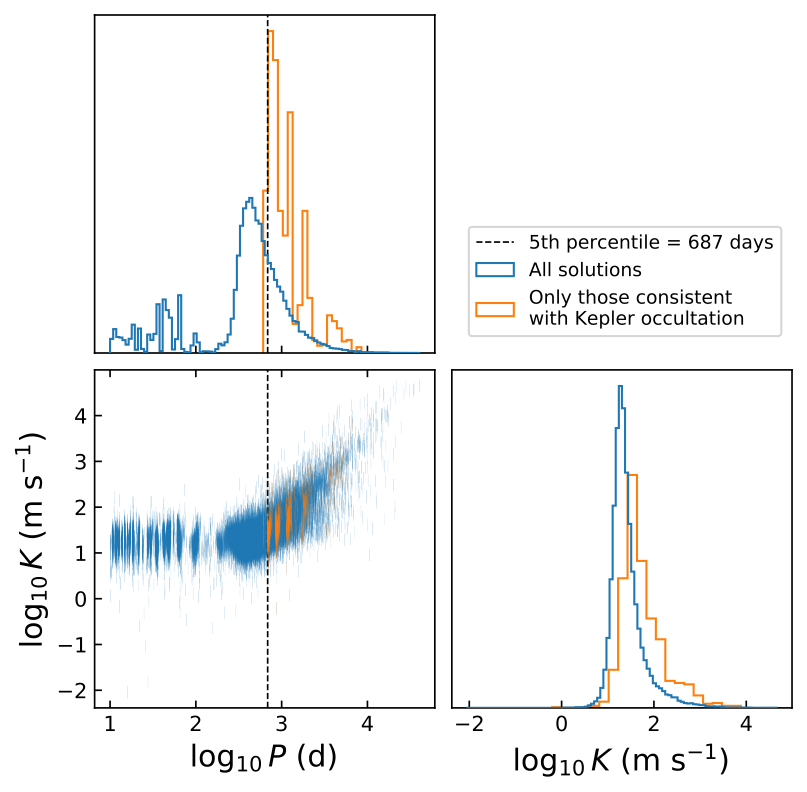}\hspace{1cm}
      \includegraphics[width=0.9\columnwidth]{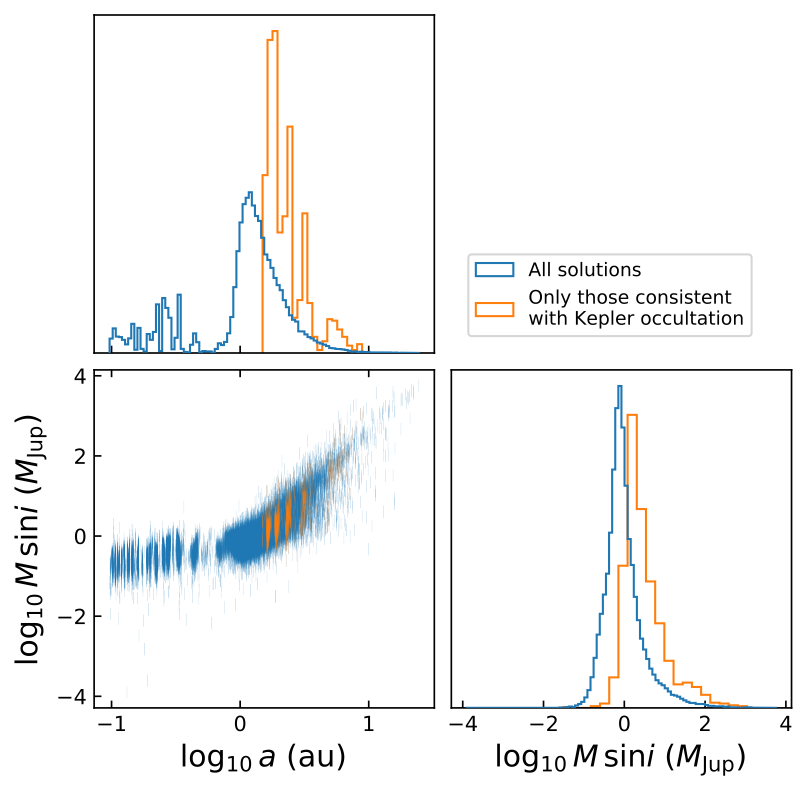}  
    \end{tabular}
  \end{center}
  \caption{Left panels: marginal posterior PDFs for orbital period ($P$) and RV semiamplitude ($K$) from the rejection sampling of the potential planetary signal. The dashed line shows the 5th percentile in orbital period for solutions that are consistent with the \Kepler\ occultation (orange). Right panels: marginal posterior PDFs for the derived parameters companion minimum mass ($M\sin{i}$) and orbital semi-major axis ($a$). Both groups of panels show the posteriors for all solutions (blue), as well as the subset consistent with the \Kepler\ occultation (orange). For the latter, 95\% of solutions have $M\sin{i}=0.6-82 \; M_{\rm Jup}$ and $P>687$~days, suggesting that the companion would likely be a long-period giant planet or brown dwarf.}
  \label{fig:planet_post}
\end{figure*}

The marginal posterior PDFs from all solutions for $P$ and $K$ are shown in Figure \ref{fig:planet_post}. The PDF for $K$ is unimodal, although non-Gaussian, and 93\% of samples are below 100~m~s$^{-1}$. The PDF for $P$ is more complicated, with multiple regions of posterior probability for orbital periods between 10 and 100~days along with a wide, non-Gaussian distribution peaked at $\sim$430~days. Almost no solutions have $P>10,000$~days. 

We also derive and display in Figure \ref{fig:planet_post} the posterior PDFs from all solutions for orbital semi-major axis ($a$) and the companion minimum mass ($M\sin{i}$). In doing so, we solve for $M\sin{i}$ numerically and do not assume that the companion mass is negligible compared to the host mass. The PDF for $a$ mirrors that for $P$. The $M\sin{i}$ posterior PDF peaks at 0.75~$M_{\rm Jup}$ and 95\% of samples are in the range of 0.2--20~$M_{\rm Jup}$. This distribution suggests that if this RV signal is caused by a companion, then that companion is likely a giant planet or a brown dwarf. 

Could this potential giant planet or brown dwarf be the cause of the occultation observed by \Kepler? The depth of the occultation is readily consistent with a grazing transit of a 1~$R_{\rm Jup}$ object. However, we address this question in more detail by considering the subset of the rejection sampling solutions that are consistent with the \Kepler\ occultation. For each solution, we determine the time of inferior conjunction (in BJD$_{\rm TDB}$) in the vicinity of the \Kepler\ occultation. We calculate the true anomaly ($f$) during transit
\begin{equation}
    f = \frac{\pi}{2} - \omega \;,
\end{equation}

\noindent the corresponding eccentric anomaly \citep[$E$;][p. 33]{Murray1999}
\begin{equation}
    E = 2 \; \mathrm{tan}^{-1} \left [ \sqrt{\frac{1-e}{1+e}} \; \tan{\left (\frac{f}{2} \right )} \right ] \;,
\end{equation}

\noindent and the corresponding mean anomaly \citep[$M$;][p. 34]{Murray1999}
\begin{equation}
    M = E - e\sin{E} \;.
\end{equation}

\noindent Finally, substituting Equation (3) from \citet{PriceWhelan2017}, we solve for the time of inferior conjunction $t_{\rm c}$
\begin{equation}
    t_{\rm c} = \frac{P}{2\pi}(\phi_{\rm p} + M) + c
\end{equation}

\noindent where $c$ is a temporal offset that shifts $t_{\rm c}$ to BJD\footnote{In v0.3 of \texttt{The Joker}, the offset $c$ equals the BJD of the first (earliest) data point.}.

We consider a solution to be consistent with the \Kepler\ occultation if its conjunction time is within $\pm$5\% of that solution's orbital period of the \Kepler\ occultation. This buffer does not reflect the precision on the measured occultation timing, but rather the limit of an individual solution's ability to represent its local region of parameter space. 

In addition to $t_c$, we also impose that solutions consistent with the \Kepler\ occultation must have $P>609$~days, which is the lower limit calculated from the \kic\ light curves in Section \ref{sec:Plim}. 

We present the posterior PDFs from only those solutions that are consistent with the \Kepler\ occultation in Figure \ref{fig:planet_post}. The histograms in this figure have each been normalized such that they integrate to unity. Relative to all solutions, those that are consistent with the occultation have longer orbital periods, larger RV semiamplitudes, and higher masses in general. Specifically, 95\% of these solutions have $P>687$~days and $M\sin{i}$ in the range of 0.6--82~$M_{\rm Jup}$. This means that if the RV signal shown in the bottom panel of Figure \ref{fig:rv} and the \Kepler\ occultation are caused by the same companion, then that companion is likely a long-period giant planet or brown dwarf.  

We further inform our interpretation of these data by considering the occultation duration. Assuming that the occulting companion has a radius of 1~$R_{\rm Jup}$, we calculate the occultation duration as a function of impact parameter ($b$). The transit duration for an eccentric orbit is approximated by \citep{Winn2010}
\begin{equation}\label{eq:T}
    T = \frac{P\sqrt{1-e^2}}{\pi(1+e\sin{\omega})}\sin^{-1} \left [\frac{R_{\star}}{a} \frac{\sqrt{(1+k)^2 - b^2}}{\sin{i}} \right ]
\end{equation}

\noindent where $k$ is the ratio of the planet radius to the stellar radius. The impact parameter can be substituted for orbital inclination ($i$) according to 
\begin{equation}
    b = \frac{a \, \cos{i}}{R_{\star}}\left ( \frac{1 - e^2}{1+e\, \sin{\omega}} \right ) \; .
\end{equation}

\begin{figure}
    \centering
    \includegraphics[width=\columnwidth]{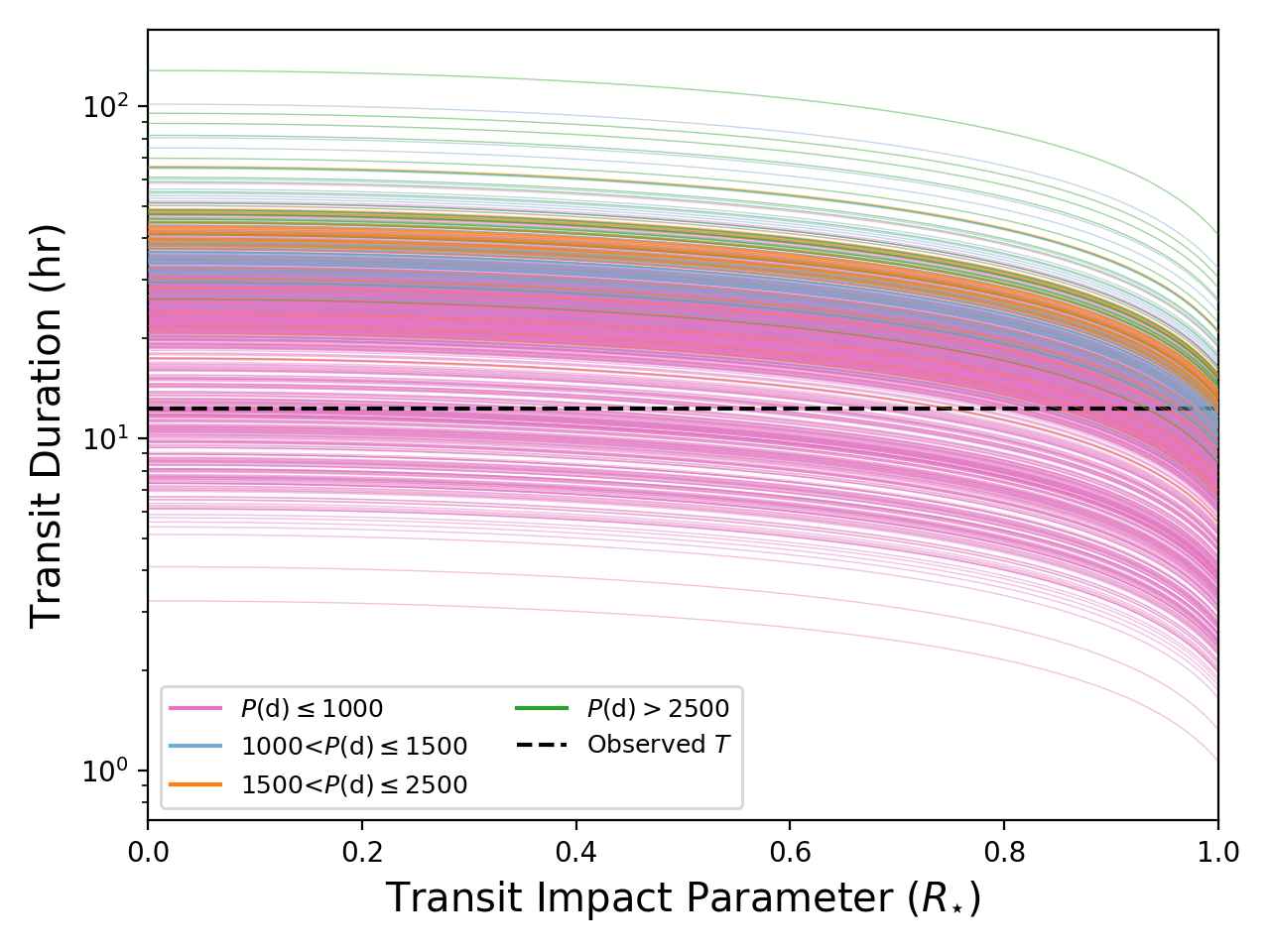}
    \caption{Transit duration ($T$) as a function of impact parameter for the solutions to the potential planetary RV signal that are consistent with the \Kepler\ occultation. The duration of the occultation observed by \Kepler\ (Figure \ref{fig:transit}) is indicated. We find that 1~$R_{\rm Jup}$ objects that have solutions with $P>1500$~days are incapable of producing the observed occultation regardless of the impact parameter.}
    \label{fig:planet_T}
\end{figure}

We calculate $T(b)$ for all solutions consistent with the \Kepler\ occultation and display the result in Figure \ref{fig:planet_T}. The duration of the observed occultation event is 12.3~hr. The curves in this figure are colored by orbital period. We find that 1~$R_{\rm Jup}$ objects that have solutions with $P>1500$~days are incapable of producing the observed occultation regardless of the impact parameter. \citet{Kawahara2019} found that $b=0.94^{+0.01}_{-0.02}$, which in this case allows some solutions with $1000<P(\mathrm{days})\lesssim1500$. Overall, the transit duration places an upper limit on orbital period ($\sim$1500~days) that complements the lower limit ($\sim$687~days).  

\subsubsection{Visualizing the RV Time Series}\label{sec:vis_planet}

\begin{figure}
    \centering
    \includegraphics[width=\columnwidth]{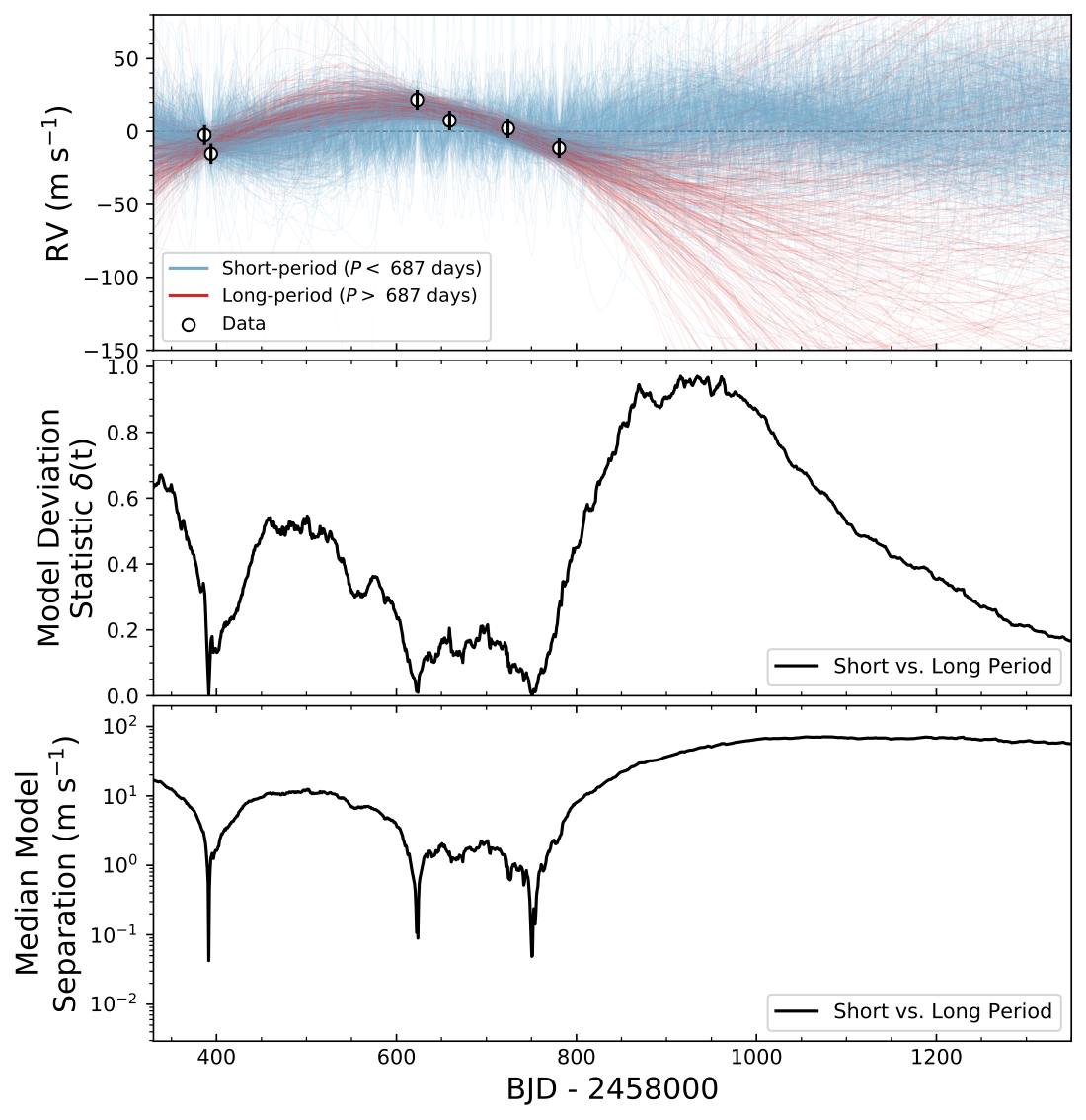}
    \caption{Top: model RV time series of the potential planetary RV signal based on 1000 random draws from the posterior PDFs of all solutions. The Keck-HIRES data are shown as open circles. Middle: the model deviation statistic time series comparing the long-period solutions (red) that are consistent with the \Kepler\ occultation and short-period solutions (blue) that are not. The most useful time to make an observation that distinguishes between groups of solutions was before BJD = 2459000. Bottom: the absolute RV separation between the median time series of each group of solutions.}
    \label{fig:rj_planet}
\end{figure}

The RV time series of the potential giant planet or brown dwarf companion can also be modeled using the posterior PDFs. We randomly draw 1000 samples from the posterior PDFs of all solutions and calculate their RV time series (Figure \ref{fig:rj_planet}, top panel). For each draw, the median value of the $\dot \gamma$ posterior is used to remove the long-term RV trend from the model time series. The median and 68\% confidence region for $\dot \gamma$ is $-7.05^{+0.06}_{-0.07}$~m~s$^{-1}$~day$^{-1}$. The same procedure is also applied to the Keck-HIRES data. We use the lower limit of the orbital period (687~days; see Figure \ref{fig:planet_post}) to distinguish between long-period solutions that are consistent with the \Kepler\ occultation and short-period solutions that are not.

To aid in predicting future observations that distinguish between groups of solutions, we define a time-dependent model deviation statistic $\delta$

\begin{equation}\label{eq:model_dev}
    \delta(t) = \frac{|M_{\rm short}(t) - M_{\rm long}(t)|}{\sqrt{\sigma_{\rm short}(t)^2 + \sigma_{\rm long}(t)^2}}
\end{equation}

\noindent where $M_{\rm short}$ and $M_{\rm long}$ are the median RV time series models of the short- and long-period groups of solutions, respectively, $\sigma_{\rm short}$ and $\sigma_{\rm long}$ are the standard deviation time series of the RV models within the short- and long-period groups of solutions, and $t$ is time. This statistic is the difference between two models weighted by the combined spread within each of those models. It is effectively a S/N ratio for the deviation between models. Values of $\delta(t)$ greater than unity highlight strategic times to obtain observations that distinguish between groups of solutions. 

As shown in Figure \ref{fig:rj_planet}, $\delta \approx 1$ between BJD = 2458900 and BJD = 2459000. After this brief period, which has already passed, the short- and long-period model groups mix such that $\delta$ remains below unity for the near-term future. In general, this means that a single observation will likely not distinguish between model groups or, by extension, constrain the nature of the companion that caused the \Kepler\ occultation. The bottom panel of Figure \ref{fig:rj_planet} displays the absolute velocity separation between the median time series of the short- and long-period model groups. This suggests that the level of precision yielded by the matched-template technique is sufficient for future observations of this target.

\subsection{The Long-term RV Trend}\label{sec:rj_eb}

We now investigate the long-term RV trend (Figure \ref{fig:rv}, top panel) with another rejection sampling analysis. Here, we do not include a parameter for first-order acceleration. Instead, the data are treated as if they cover only a fraction of the phase of a longer signal. We effectively mask the potential planetary signal by increasing the error bars of the Keck-HIRES data by a factor of four. This increase makes the new median error (of $\sim$27~m~s$^{-1}$) consistent with the peak of the potentially planetary semiamplitude posterior PDF (Figure \ref{fig:planet_post}). If we do not increase the data error, the rejection sampling cannot robustly sample the posterior PDFs of the model parameters merely because the model likelihood for all samples is low.  

\begin{figure}
    \centering
    \includegraphics[width=\columnwidth]{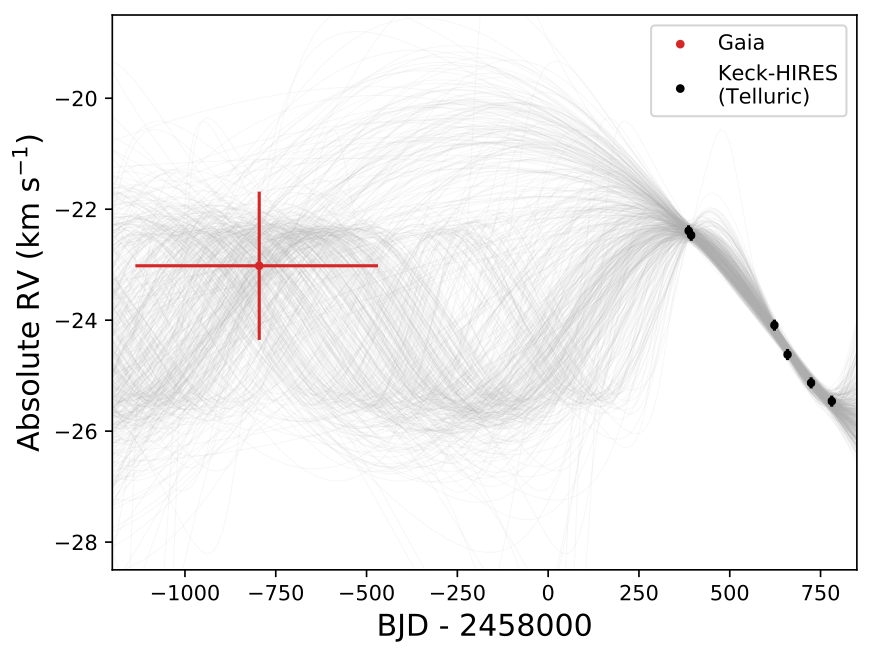}
    \caption{Absolute, telluric-calibrated RVs from Keck-HIRES (black points) and median RV measurement spanning BJD = 2456863.5--2457531.5 from \Gaia\ (red point). The \Gaia\ measurement shows that the linear trend of the Keck-HIRES data must turn over, thereby setting an upper limit on the orbital period of the companion causing the trend ($P_{\rm max} \approx 13,000$~days). The gray lines are 500 time series created by randomly drawing solutions from a rejection sampling analysis of these RVs.}
    \label{fig:gaia}
\end{figure}

For this rejection sampling analysis, the lower bound on the orbital period prior is set to the observational baseline ($P_{\rm min}=394$~days) since the data clearly do not span a full orbit. The choice of an upper bound for this prior is less straightforward. The high degree of linearity of the precise Keck-HIRES RVs is consistent with a broad range of long periods. However, we can constrain the RV variation of \kic\ with its RV measurements from the \Gaia\ mission \citep{Gaia2018,Katz2019}. Between 2014 July 25 (BJD = 2456863.5) and 2016 May 23 (BJD = 2457531.5), \Gaia\ acquired 11 absolute RV measurements of \kic\footnote{According to the Gaia data archive \url{https://gea.esac.esa.int/archive/}, accessed 2020 April 19.}. These individual measurements are yet unpublished, but the \Gaia\ Data Release 2 included their median value ($-23.0\pm1.3$~km~s$^{-1}$) and the corresponding epoch (2015.5 or BJD = 2457204.5). This \Gaia\ RV can be compared to the telluric-calibrated, absolute RVs produced from the HIRES spectra by the methodology of \citet{Chubak2012}, which we list in Table \ref{tab:rv}. Using standard stars, \citet{Soubiran2018} found excellent agreement between the \Gaia\ RVs and the catalog of \citet{Chubak2012}, quantified by a dispersion of only 0.072~km~s$^{-1}$, which is much lower than the error on the \Gaia\ RV measurement of \kic. In Figure \ref{fig:gaia}, we show the \Gaia\ RV and the Keck-HIRES telluric RVs of \kic. It is clear that the linear trend of the Keck-HIRES data must turn over in order for these data to be consistent with the measurement from \Gaia. This suggests that we can derive an upper limit on the orbital period of the companion causing the long-term trend. 

To derive an upper limit on the orbital period of the companion causing the long-term trend, we conduct a simple rejection sampling analysis on the joint Keck-\Gaia\ RV data set using \texttt{The Joker}. We bound the orbital period prior at 394~days and 1\e{6}~days and sample the posterior $2^{21}$ times. From the resulting posterior PDF for orbital period, we calculate the 99.7th percentile to be $\sim$13,000~days. We treat this value as the maximum orbital period of the companion causing the long-term trend.  

With the orbital period prior bounded at $P_{\rm min}=394$~days and $P_{\rm max}=13,000$~days, we proceed with the rejection sampling analysis on just the precise Keck-HIRES RVs calculated using the matched-template analysis. We use \texttt{The Joker} to sample the prior PDFs $2^{21}$ times. Of these, 39,666 samples of the posterior PDFs survive. 

\begin{figure*}
  \begin{center}
    \begin{tabular}{cc}
      \includegraphics[width=0.9\columnwidth]{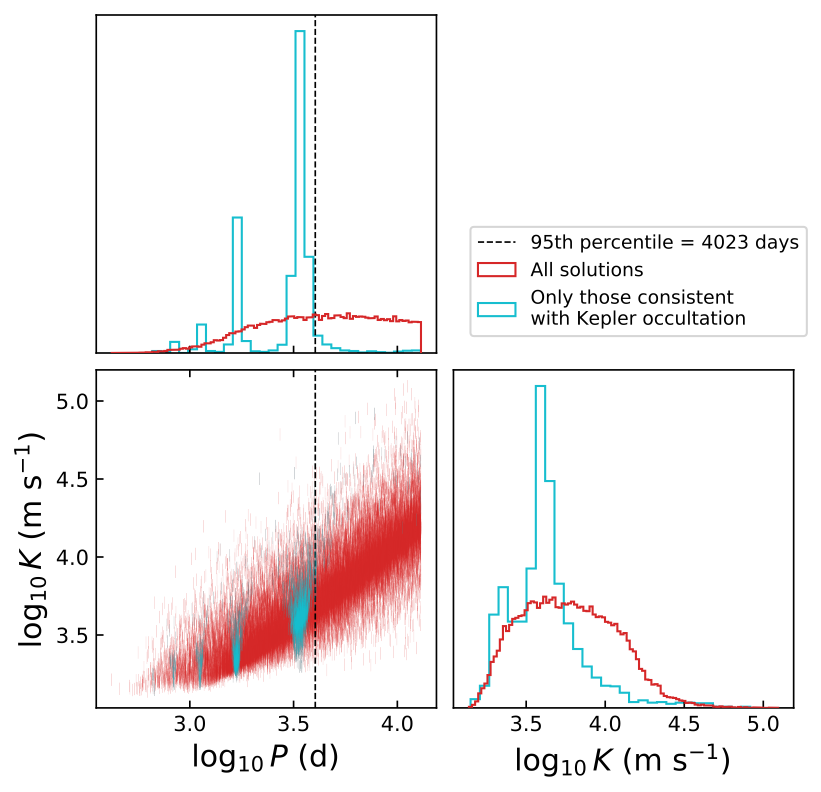} \hspace{1cm}
      \includegraphics[width=0.9\columnwidth]{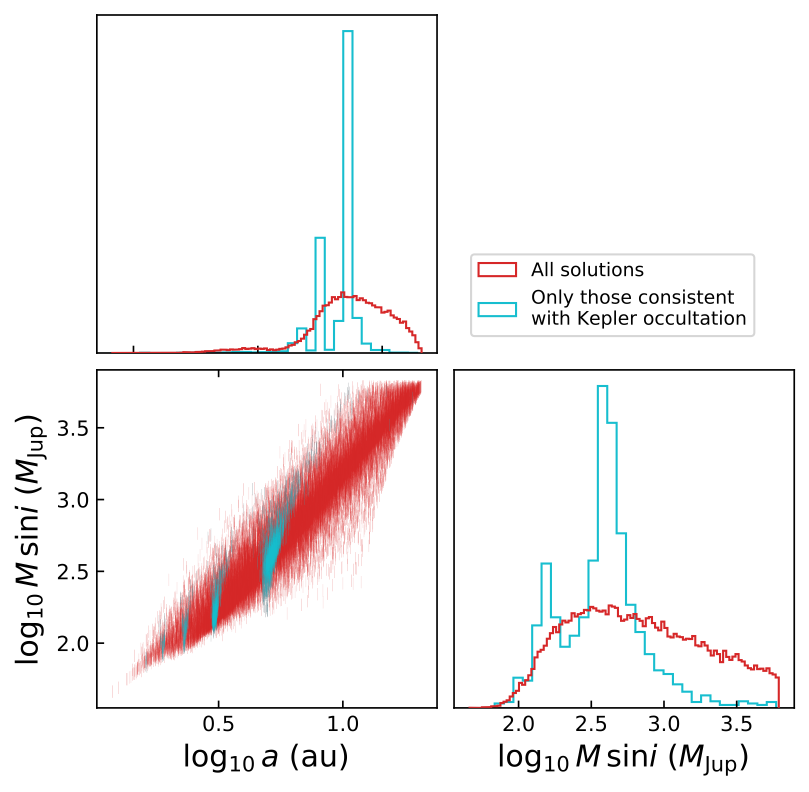}  
    \end{tabular}
  \end{center}
  \caption{Left panels: marginal posterior PDFs for orbital period ($P$) and RV semiamplitude ($K$) from the rejection sampling of the long-term RV trend. The dashed line shows the 95th percentile in orbital period for solutions that are consistent with the \Kepler\ occultation (cyan). Right panels: marginal posterior PDFs for the derived parameters companion minimum mass ($M\sin{i}$) and orbital semi-major axis ($a$). Both groups of panels show the posteriors from all solutions (red), as well as the subset consistent with the \Kepler\ occultation (cyan). For the latter, 95\% of solutions have $M\sin{i}=95-1659 \; M_{\rm Jup}$ and $P<4023$~days, suggesting that the companion would likely be a star.}
  \label{fig:star_post}
\end{figure*}

The marginal posterior PDFs for $P$ and $K$ from all solutions are shown in Figure \ref{fig:star_post}. The PDFs for $K$ and $P$ are smooth, although the latter truncates at 13,000 days owing to the prior. The shortest-period solutions are on the order of 420~days. We also derive and display the posterior PDFs from all solutions for orbital semi-major axis ($a$) and the companion minimum mass ($M\sin{i}$). In doing so, we solve for $M\sin{i}$ numerically and do not assume that the companion mass is negligible compared to the host mass. The $M\sin{i}$ posterior PDF truncates at high values as a result of the minimum possible values of $K$ set by the span of the RV data.  

As in Section \ref{sec:rj_planet}, it is informative to identify the subset of the posterior PDF that is consistent with the occultation observed by \Kepler. We again compare each solution's inferior conjunction time with the time of the \Kepler\ occultation as described in Section \ref{sec:rj_planet}. We present the posterior PDFs from only those solutions that are consistent with the \Kepler\ occultation in Figure \ref{fig:star_post}. The histograms in this figure have each been normalized such that they integrate to unity. This posterior PDF for $P$ is multimodal, with distinct peaks at $\sim1700$ and $\sim3400$-days. The 95th percentile in orbital period for solutions that are consistent with the \Kepler\ occultation is 4023~days. The posterior PDF for $M\sin{i}$ is broad and also multimodal with peaks at $\sim155$ and $\sim400$~$M_{\rm Jup}$. Both of these cases, and the vast majority of all solutions, suggest that the companion causing the long-term trend is stellar in nature.

\subsubsection{Visualizing the RV Time Series}

\begin{figure}
    \centering
    \includegraphics[width=\columnwidth]{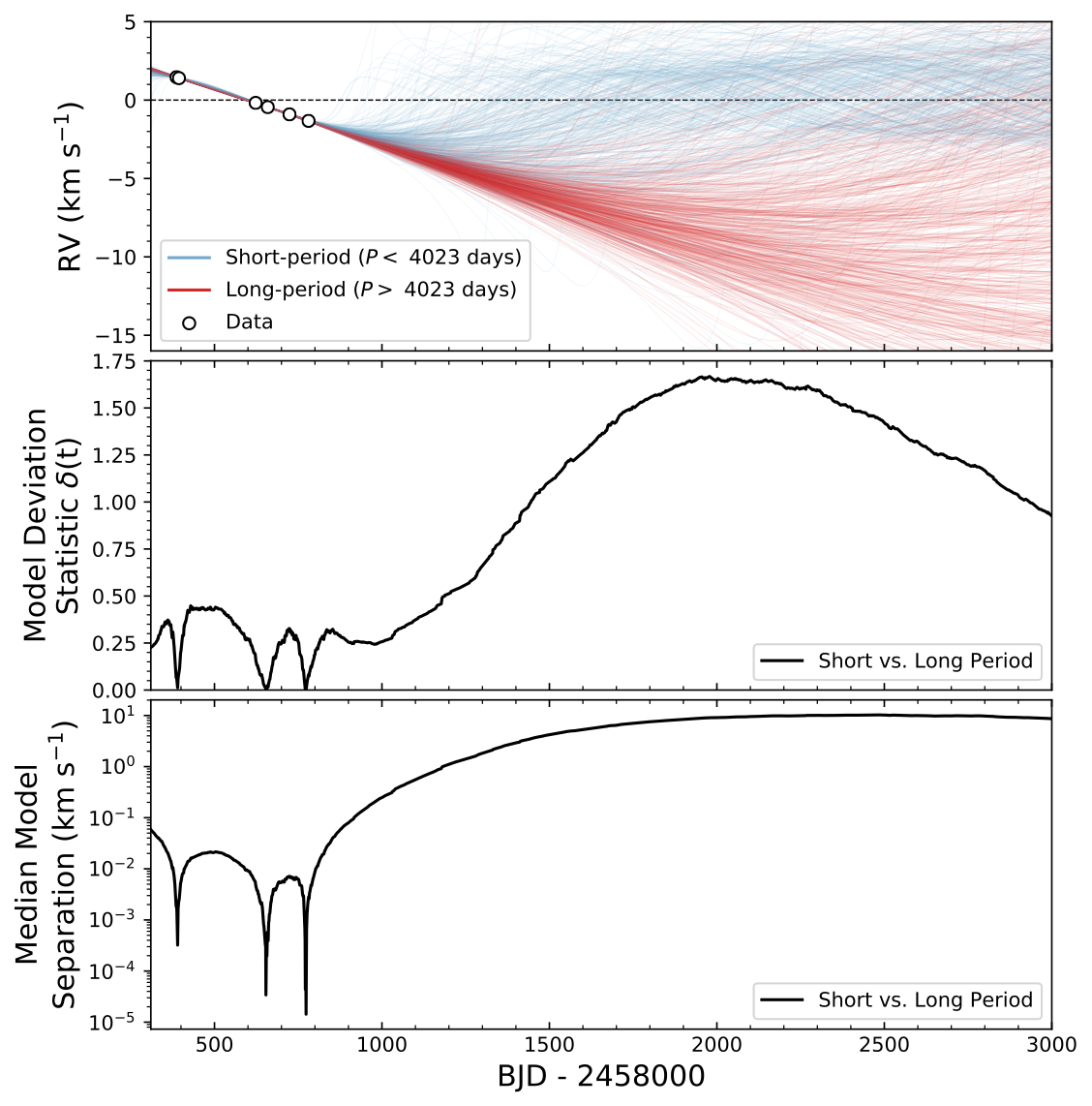}
    \caption{Top: model RV time series of the long-term RV trend for \kic\ based on 1000 random draws from the posterior PDF. The Keck-HIRES data are the open black circles. The short- and long-period solutions are divided at $P=4023$~days, which is the 95th percentile in orbital period for solutions consistent with the \Kepler\ occultation. Middle: the model deviation statistic time series (Section \ref{sec:vis_planet}). The broad peak represents a balance between the divergence of short- and long-period models and the increasing uncertainty within each model group. Bottom: The absolute RV separation between median short- and long-period models.}
    \label{fig:rj_star}
\end{figure}

We model the RV time series of the stellar companion causing the long-term trend in the RVs. We randomly draw 1000 samples from the posterior PDFs of all the solutions and calculate their RV time series (Figure \ref{fig:rj_star}, top panel). We use the upper limit of the orbital period (4023~days; see Figure \ref{fig:star_post}) to distinguish between short-period solutions that are consistent with the \Kepler\ occultation and long-period solutions that are not. From the median RV times series of short- and long-period groups of solutions, we calculate the RV separation and the model deviation statistic $\delta(t)$. As derived in Section \ref{sec:vis_planet}, $\delta(t)$ is the time-dependent difference between median models weighted by the uncertainty in that difference. The middle panel of Figure \ref{fig:rj_star} shows a broad peak in $\delta(t)$ near BJD$_{\rm TDB}=$~2460000. This peak and subsequent turnover demonstrate how the divergence between the short- and long-period models is eventually overcome by the increasing uncertainty within each group. The bottom panel of Figure \ref{fig:rj_star} shows that the RV separation between the model groups is substantial ($\sim$10~km~s$^{-1}$) near the peak in $\delta(t)$.  

The information displayed in Figure \ref{fig:rj_star} suggests that the optimal time to obtain another RV epoch of \kic\ is near BJD = 24560000 (2023 February). At that time, the short- and long-period model groups will be substantially separated such that a single, high-precision RV measurement could likely distinguish between them and, by extension, constrain the nature of the companion that caused the occultation seen by \Kepler.

\subsection{Results of the Rejection Sampling Analysis}\label{sec:rj_results} 

The rejection sampling analysis of the previous sections is meant to characterize the signals in the sparse set of RV observations and enable an informed interpretation of the single occultation event detected in Quarter 4 of the \Kepler\ primary mission. We find that there are two distinct signals in the RV data. The first signal is a several kilometers-per-second trend from what is likely a stellar companion. The posterior PDF for the minimum mass of this companion is too broad to determine its exact nature, but it is highly unlikely to be planetary. Its orbital period is likely longer than $\sim$1000~days, but not so long that we would expect it to have been detected by the AO imaging (Figure \ref{fig:nirc2}). The depth and duration of the occultation event from \Kepler\ are consistent with the properties of this companion, but only if its orbital period is less than $\sim$4023~days or likely either $\sim1700$ or $\sim$3400~days. If this companion's orbital period is substantially greater than $\sim$4023~days, then only very fine-tuned combinations of orbital parameters would have allowed it to have passed through inferior conjunction during Quarter 4 of the primary \Kepler\ mission. 

Using the RV time series and the model deviation statistic in Figure \ref{fig:rj_star}, we predict that just one additional RV epoch acquired in the 2022 or 2023 \Kepler\ observing seasons (surrounding 2460000~BJD) would be particularly helpful in the interpretation of the \kic\ system. At that time, the short-period solutions (which are consistent with the occultation) and the long-period solutions (which are not) will have sufficiently diverged to be able to distinguish between the two. If this RV observation follows the long-period solutions (i.e., the red curves in Figure \ref{fig:rj_star}), then the \Kepler\ occultation was likely not caused by the stellar companion. If the RV observation lands in between the two groups or toward the short-period groups, then the stellar companion still may have caused the occultation event. In this case, many additional observations will be necessary to characterize the system. 

The second signal in the RV data is that of a potential giant planet or brown dwarf. This signal is seen when the large linear trend is removed (Figure \ref{fig:rv}, bottom panel). There are several somewhat discrete groups of possible orbital periods for this companion, many of which are shorter than 1000~days. A total of 88\% of the solutions qualify this companion as a giant planet with minimum mass in the range of 0.3--13~$M_{\rm Jup}$. Assuming a typical 1~$R_{\rm Jup}$ radius, this planetary or brown dwarf companion is capable of producing the occultation event observed by \Kepler. Also, the fact that this companion likely orbits closer to \kic\ than the stellar companion means that it is geometrically more likely to have caused the occultation. However, based on the date of the occultation, its duration, and the nondetection of a second event in the full \Kepler\ data set, the planetary or brown dwarf companion must have an orbital period in the range of $\sim$687--1500~days to have caused the occultation event. This suggests that it would be worthwhile to obtain several RVs to distinguish between various long-period solutions (as described in Section \ref{sec:rj_planet}) and then conduct follow-up photometric monitoring to detect an additional transit. 

%%%%%%%%%%%%%%%%%%%%%%%%%%%%%%%%%%%%%%%%%%%%%%%%%%%%%%%%%%%%%%%%%%%%%%%%%%

\section{Discussion}\label{sec:discussion}

We found that the story of \kicb---a member of the confirmed exoplanets list---is more complicated than previously thought. Although we cannot uniquely determine the full nature and architecture of the \kic\ system, we know that the current description of \kicb\ as it stands in the list of confirmed exoplanets is incorrect. \kic\ likely hosts a substellar-mass companion with an orbital period less than or around a few thousand days. It could be a giant planet or a brown dwarf, and its radius is unknown because it may not transit its host star.  

Despite the uncertainty pertaining to the nature and architecture of the \kic\ system, the scientific potential surrounding the future characterization of this system remains high. Regardless of which object caused the occultation event observed by \Kepler, \kic\ appears to be a binary star system where the primary hosts a giant planet or brown dwarf. If we assume the former, this system would join a relatively small group of circumprimary exoplanets, which are useful for probing the extremes of planet formation, as well as for comparison to single-star planetary systems \citep[e.g.,][and references therein]{Thebault2015}. Even within the sample of known circumprimary exoplanets, \kicb\ would be quite interesting. To date, only three systems (Kepler-420, \citep{Santerne2014}; Kepler-693, \citep{Masuda2017}; and HD~42936, \citep{Barnes2019}) are known to have a circumprimary planet and a secondary star within 10 au\footnote{Based on Figure 1 from \citet{Thebault2015}, which is also maintained at \url{http://exoplanet.eu/planets_binary/} and updated as of 2020 February 2.}. 

The \kic\ system becomes even more interesting when we consider which companion caused the \Kepler\ occultation event. If the stellar companion caused the occultation, then \kic\ joins the list of known eclipsing binaries (EBs). Assuming that the binary orbital period is $\sim$3400~days (the highest peak in Figure \ref{fig:star_post}), \kic\ would rank in the 99.9th percentile by period among other EBs \citep{Malkov2006}. According to the \Kepler\ Eclipsing Binary Catalog \citep[e.g.,][]{Prsa2011}\footnote{\url{http://keplerebs.villanova.edu/}, updated 2019 August 8.}, \kic\ would become the longest-period \Kepler\ EB if its period were 3400~days or even the alias at half of that value. With additional refinement of the binary star ephemeris, future eclipse observations would be particularly useful to characterizing this system. 

Alternatively, the scenario in which the giant planet or brown dwarf companion caused the occultation event is perhaps even more tantalizing. To be consistent with the impact parameter measured by \citet{Kawahara2019}, this object likely has an orbital period between 687 and 1500~days. This alone would make it a remarkable exoplanet or brown dwarf, since the geometric bias of the transit method so severely limits detections to short-period objects. Long-period transiting exoplanets are a valuable pathway toward characterizing the atmospheres, interiors, and formation histories of cooler planets that more resemble the solar system \citep[e.g.,][]{Dalba2015,Dalba2019c}. Conducting future photometric observations to detect additional occultations would be useful for measuring the bulk density of \kicb, be it a planet or brown dwarf, and for drawing further conclusions on its formation and evolution.

Finally, the evolving nature of the narrative describing the \kic\ system is one that will become more common due to the ongoing transit hunting efforts of \TESS\ \citep{Ricker2015}. \TESS\ is predicted to discover on the order of 1000 single-transit events in its primary mission alone \citep{Villanueva2019,Dalba2020a,Eisner2020,Gill2020b}. The need to conduct imaging, photometric, and spectroscopic follow-up for many of these discoveries will place an immense burden on the pool of observational resources available to the exoplanet community. 

In this work, we have utilized archival data and collected only a small amount (i.e., six RV epochs) of new follow-up data of \kic. We then conducted an exploratory study of the degeneracies between interpretations of the system's nature and architecture. This has led to predictions for future observations of this system. An alternate approach would have been to acquire as many RV observations as possible over a full orbit. This approach is commonly applied in exoplanet characterization endeavors, although most transiting exoplanets followed up with RVs have much shorter orbital periods than the companions orbiting \kic. Moving forward, we argue that the conservative approach---whereby degeneracies in the interpretation of the system are identified and the timing of efficient follow-up opportunities is determined---will become increasingly valuable for the characterization of single-transit planet candidates. The deluge of \TESS\ short-period planet discoveries places a strain on existing RV facilities. Attempts to observe single-transit planet candidates---one of the only avenues to long-period exoplanets suitable for detailed characterization---must find a way to complement efforts to observe shorter-period planets. Our conservative approach is one option, and we have demonstrated its usefulness for the \kic\ system in this work.

%%%%%%%%%%%%%%%%%%%%%%%%%%%%%%%%%%%%%%%%%%%%%%%%%%%%%%%%%%%%%%%%%%%%%%%%%%

\section{Summary}\label{sec:summary}

We described two novel techniques surrounding the analysis of precise RVs of a supposed exoplanet hosting system \kic. The first technique pertains to the extraction of the RVs of \kic, a 13th-magnitude F5 star previously observed by the primary \Kepler\ mission. To extract precise RVs for this star, we developed a novel matched-template technique that leverages the collection of several hundred high-S/N iodine-free spectra acquired with Keck-HIRES (Section \ref{sec:match_temp}). This technique matches a new target star to a current member of the template library and uses its preexisting template to extract precise RVs. Using this method, we were able to forgo the collection of a new iodine-free template spectrum for \kic, which would have required an expensive investment of time. We found that the RV uncertainty incurred by the matched-template technique (in addition to internal RV errors) is 4--8~m~s$^{-1}$ for most stars that would be subject to precise RV observations (Figure \ref{fig:resid_rms}). In Section \ref{sec:validate}, we provide a procedure for determining the suitable error to add to the internal RVs for different types of stars. In general, the matched-template technique will produce RVs that are precise and accurate enough to investigate giant planet signals and to aid in the planning of additional observations.

The second novel technique described in this work surrounds the analysis of a single-transit exoplanet candidate through the synthesis of sparse collections of imaging, photometric, and spectroscopic data. \kic\ was previously thought to host a long-period transiting exoplanet based on a single occultation event detected in Quarter 4 of the \Kepler\ primary mission \citep{Wang2015,Kawahara2019}. We used the rejection sampling tool \texttt{The Joker} \citep{PriceWhelan2017} to model the RV observations of \kic\ and explore the possible explanations for the nature of the system. We found that the published parameters for the confirmed planet \kicb\ are incorrect and that the \kic\ system is substantially more complicated than originally thought. 

The RVs of \kic\ show a large trend that is indicative of a stellar companion with an orbital period longer than a few thousand days. In addition to the RV trend, there also exists a signal likely caused by a giant planet or brown dwarf companion with mass in the range of 0.6--82~$M_{\rm Jup}$ and orbital period less than a few thousand days. Based on these findings, we cannot clearly identify which companion to \kic\ caused the occultation event detected by \Kepler. If the occultation was caused by the giant planet or brown dwarf, then we can further constrain its orbital period to be between $\sim$687 and $\sim$1500~days. Alternatively, if the stellar companion caused the occultation, we can further constrain its orbital period to be less than 4023~days, and likely either $1700$ or $3400$~days. 

We offer predictions to be tested by future observations that can distinguish between possible scenarios for the architecture of the \kic\ system. Regardless of which scenario is correct, we explain why \kic\ is a rare and unlikely system worthy of follow-up characterization. Lastly, we argue that investigations conducted after only a small amount of follow-up data have been collected will be critical to maximizing the science return from single-transit objects detected by \Kepler\ and \TESS.  

%%%%%%%%%%%%%%%%%%%%%%%%%%%%%%%%%%%%%%%%%%%%%%%%%%%%%%%%%%%%%%%%%%%%%%%%%%

\acknowledgements

The authors thank the anonymous referee for thoughtful comments that improved the quality and clarity of this work. The authors acknowledge Erik Petigura for having the idea to use previously acquired templates rather than obtaining new ones for faint stars. The authors also thank Lee Rosenthal for helpful conversations about \texttt{The Joker}. Lastly, the authors acknowledge all of the observers on the California Planet Search team for their many hours of hard work. P.D. is supported by a National Science Foundation (NSF) Astronomy and Astrophysics Postdoctoral Fellowship under award AST-1903811.

This research has made use of the NASA Exoplanet Archive, which is operated by the California Institute of Technology, under contract with the National Aeronautics and Space Administration under the Exoplanet Exploration Program. This paper includes data collected by the \Kepler\ mission and obtained from the MAST data archive at the Space Telescope Science Institute (STScI). Funding for the Kepler mission is provided by the NASA Science Mission Directorate. STScI is operated by the Association of Universities for Research in Astronomy, Inc., under NASA contract NAS 5–26555. This work presents results from the European Space Agency (ESA) space mission \Gaia. Gaia data are being processed by the \Gaia Data Processing and Analysis Consortium (DPAC). Funding for the DPAC is provided by national institutions, in particular the institutions participating in the Gaia MultiLateral Agreement (MLA).

Some of the data presented herein were obtained at the W. M. Keck Observatory, which is operated as a scientific partnership among the California Institute of Technology, the University of California, and NASA. The Observatory was made possible by the generous financial support of the W.M. Keck Foundation. Finally, the authors wish to recognize and acknowledge the very significant cultural role and reverence that the summit of Maunakea has always had within the indigenous Hawaiian community. We are most fortunate to have the opportunity to conduct observations from this mountain.

\vspace{5mm}
\facilities{Keck:I(HIRES), Keck:II(NIRC2), Kepler, Gaia}

\vspace{5mm}
\software{\texttt{Lightkurve} \citep{Lightkurve2018}, \texttt{SpecMatch} \citep{Petigura2015,Petigura2017}, \texttt{SpecMatch-Emp} \citep{Yee2017}, \texttt{The Joker} \citep{PriceWhelan2017}}

\end{document}